\documentclass[a4paper,11pt]{article}
\pdfoutput=1 

\usepackage{jcappub} 
\usepackage{indentfirst}
\usepackage{appendix}
\usepackage{multirow}

\usepackage{array}
\usepackage{enumerate}
\usepackage{graphicx}
\usepackage{dcolumn}
\usepackage{bm}
\usepackage{amssymb}
\usepackage{latexsym}
\usepackage{booktabs}
\usepackage{amsmath}
\usepackage{url}

\begin{document}
\title{Standard siren cosmology in the era of the 2.5-generation ground-based gravitational wave detectors: bright and dark sirens of LIGO Voyager and NEMO}

\author[a]{Shang-Jie Jin,}
\author[a]{Rui-Qi Zhu,}
\author[a]{Ji-Yu Song,}
\author[a]{Tao Han,}
\author[a]{Jing-Fei Zhang}
\author[a,b,c,1]{and Xin Zhang\note{Corresponding author.}}
\affiliation[a]{Key Laboratory of Cosmology and Astrophysics (Liaoning Province) and College of Sciences, Northeastern University, Shenyang 110819, China}
\affiliation[b]{Key Laboratory of Data Analytics and Optimization for Smart Industry (Ministry of Education), Northeastern University, Shenyang 110819, China}
\affiliation[c]{National Frontiers Science Center for Industrial Intelligence and Systems Optimization, Northeastern University, Shenyang 110819, China}

\emailAdd{jinshangjie@stumail.neu.edu.cn, zhuruiqi@stumail.neu.edu.cn, songjiyu@stumail.neu.edu.cn, hantao@stumail.neu.edu.cn, jfzhang@mail.neu.edu.cn, zhangxin@mail.neu.edu.cn}

\abstract{The 2.5-generation (2.5G) ground-based gravitational wave (GW) detectors LIGO Voyager and NEMO are expected to be operational in the late 2020s and early 2030s. In this work, we explore the potential of GW standard sirens observed by the 2.5G GW detectors in measuring cosmological parameters, especially for the Hubble constant. Using GWs to measure cosmological parameters is inherently challenging, especially for 2.5G detectors, given their limited capability, which results in weaker constraints on cosmological parameters from the detected standard sirens. However, the measurement of the Hubble constant using standard siren observations from Voyager and NEMO is still promising. For example, using bright sirens from Voyager and NEMO can measure the Hubble constant with a precision of about $2\%$ and $6\%$ respectively, and using the Voyager-NEMO network can improve the precision to about $1.6\%$. Moreover, bright sirens can be used to break the degeneracy of cosmological parameters generated by CMB data, and to a certain extent, 2.5G detectors can also play a role in this aspect. Observations of dark sirens by 2.5G detectors can achieve relatively good results in measuring the Hubble constant, with a precision of within $2\%$, and if combining observations of bright and dark sirens, the precision of the Hubble constant measurement can reach about $1.4\%$. Finally, we also discussed the impact of the uncertainty in the binary neutron star merger rate on the estimation of cosmological parameters. We conclude that the magnificent prospect for solving the Hubble tension is worth expecting in the era of the 2.5G ground-based GW detectors.}


\maketitle
\section{Introduction}
In recent decades, the study of cosmology has achieved great success. The $\Lambda$ cold dark matter ($\Lambda$CDM) model, which is now viewed as the standard model of cosmology, has been established and strongly favors the current observations. In particular, based on the standard $\Lambda$CDM model, the precise measurements of cosmic microwave background (CMB) anisotropies have ushered in the era of precision cosmology \cite{WMAP:2003elm,WMAP:2003ivt}. However, in recent years, with the improvements of the measurement precisions of cosmological parameters, some puzzling tensions appear. Notably, the tension between the values of the Hubble constant inferred from the Planck CMB observations (0.8\% precision, assuming the $\Lambda$CDM model) \cite{Planck:2018vyg} and obtained through distance ladder measurements (1.4\% measurement, model-independent) \cite{Riess:2021jrx} has now reached more than $5\sigma$ \cite{Riess:2021jrx}. Recently, the Hubble tension has been widely discussed in the literature \cite{Verde:2019ivm,Riess:2019qba,DiValentino:2021izs,Kamionkowski:2022pkx,Perivolaropoulos:2021jda,Abdalla:2022yfr,Guo:2018ans,Poulin:2018cxd,Cai:2021wgv,Gao:2021xnk,Yang:2018euj,DiValentino:2019jae,DiValentino:2020zio,Liu:2019awo,Zhang:2019cww,Ding:2019mmw,Li:2020tds,Vagnozzi:2021tjv,Guo:2019dui,Feng:2019jqa,Gao:2022ahg,Cao:2021zpf}. The Hubble tension may imply the possibility of new physics beyond the standard model of cosmology. However, there is currently no consensus on a definitive extended cosmological model that can effectively resolve the Hubble tension. On the other hand, it is also important to develop late-universe cosmological probes that can independently measure the Hubble constant to make an arbitration for the Hubble tension. The gravitational wave (GW) standard siren method is one of the most promising methods.

The absolute luminosity distance can be directly obtained through the analysis of the GW waveform, which is called a standard siren \cite{Schutz:1986gp,Holz:2005df}. If the redshift information could also be obtained by identifying electromagnetic (EM) counterparts (this type of standard sirens are typically referred to as ``bright sirens'') or statistically from cross-matching of GW and galaxy catalogs (this type of standard sirens are typically referred to as ``dark sirens''), the distance-redshift relation could be used to explore the expansion history of the universe, which is widely discussed in the literature \cite{Nissanke:2009kt,Dalal:2006qt,Chen:2017rfc,Cutler:2009qv,Cai:2016sby,Cai:2017cbj,Gray:2019ksv,Vitale:2018wlg,Lagos:2019kds,Camera:2013xfa,Mukherjee:2019qmm,DAgostino:2019hvh,Wang:2018lun,Cai:2017aea,Jin:2020hmc,Belgacem:2019tbw,Howlett:2019mdh,Zhang:2019loq,Wu:2022dgy,Ezquiaga:2022zkx,Wang:2019tto,Zhao:2019gyk,Wang:2021srv,Jin:2021pcv,Yu:2021nvx,Chen:2023dgw,LISACosmologyWorkingGroup:2022jok,Dhani:2022ulg,Zhu:2021bpp,Qi:2021iic,Jin:2022tdf,Wang:2022oou,Song:2022siz,Jin:2023zhi,Hou:2022rvk,Borhanian:2020vyr,Jin:2023sfc,Jin:2022qnj,Zhu:2023jti,Han:2023exn,Li:2023gtu,Muttoni:2023prw,Branchesi:2023mws,Zheng:2024mbo,Dong:2024bvw,Zheng:2022gfi,Feng:2024lzh,Yu:2023ico,Xiao:2024nmi}.
The landmark event of the first detected binary neutron star (BNS) merger event, GW170817, together with its EM counterparts, initiated the era of multi-messenger astronomy \cite{LIGOScientific:2017vwq,LIGOScientific:2017ync}. The only bright siren event has given the first measurement of the Hubble constant with a precision of 14\% \cite{LIGOScientific:2017adf}.
For the dark sirens, the latest constraint precision of the Hubble constant given by LIGO-Virgo-KAGRA is 19\% \cite{LIGOScientific:2021aug}. Neither method is sufficient to resolve the Hubble tension conclusively. Consequently, future observations and next-generation observatories will be crucial in shedding light on solving the Hubble tension.

The 2.5-generation (2.5G) GW observatories, LIGO Voyager \cite{LIGO:2020xsf} (abbreviated as Voyager hereafter) and Neutron Star Extreme Matter Observatory (NEMO) \cite{Ackley:2020atn}, will begin observing in the late 2020s and early 2030s. Voyager and NEMO would improve the detection sensitivity by a factor of about 5 over the current aLIGO, which can greatly improve the detection rates of compact binary coalescences \cite{Evans:2021gyd}.
{Furthermore, Voyager and NEMO are expected to form a detector network, which can improve the ability to localize GW sources.}
Therefore, the standard siren observations in the era of the 2.5G GW detectors may play an important role in cosmological parameter estimations, especially for measuring the Hubble constant.

{Recently, GW astronomy in the 2.5G GW detector era was discussed in, e.g., refs.~\cite{Sarin:2021qqo,Evans:2021gyd,Ballmer:2022uxx,Relton:2021cax,Borhanian:2022czq}.
However, a detailed analysis of the GW standard siren observations from the 2.5G GW detectors in the cosmological parameter estimations is still absent and needs to be further discussed.
Therefore, in this work, we wish to investigate the potential of the 2.5G GW detectors by considering three detection strategies: Voyager, NEMO, and the Voyager-NEMO network, to address three key questions}: (i) what precisions the cosmological parameters could be measured using the bright sirens of 2.5G GW detectors, (ii) what role the bright sirens of 2.5G GW detectors could play in helping break the cosmological parameter degeneracies generated by the EM observations, and (iii) what precision the Hubble constant could be measured using the dark sirens of 2.5G GW detectors. Through our analysis, we aim to shed light on these aspects and explore the potential role of 2.5G GW detectors in helping solve the Hubble tension and advancing our understanding of cosmology.

This work is organized as follows. In section~\ref{sec2}, we introduce the methodology of simulating bright and dark siren data. In section~\ref{sec3}, we give the constraint results and make some relevant discussions. The conclusion is given in section~\ref{sec4}. {Throughout this work, we adopt the flat $\Lambda$CDM model as the fiducial model to generate the mock standard siren data with $\Omega_{\rm m}=0.3166$ and $H_0=67.27~\rm km~s^{-1}~Mpc^{-1}$ from the constraint results of Planck 2018 TT,TE,EE+lowE \cite{Planck:2018vyg}.}

\section{Methodology}\label{sec2}

\subsection{Simulations of GW events}\label{sec2.1}
In the observation frame, the merger rate as a function of the redshift is given by
\begin{align}
   R_{\rm obs}(z)=\frac{R_{\rm m}(z)}{1+z}\frac{dV(z)}{dz},
   \label{eq:rate1}
\end{align}
where the $(1+z)$ term arises from converting the source-frame time to the observation-frame time and $dV/dz$ is the comoving volume element. $R_{\rm m}$ is the source-frame merger rate, which is related to the cosmic star formation rate,
\begin{align}
R_{\rm m}(z_{\rm m}) =\int^{\infty}_{z_{\rm m}} dz_{\rm f} \frac{dt_{\rm f}}{dz_{\rm f}} R_{\rm sf} (z_{\rm f}) P(t_{\rm d}),
\label{eq:rate2}
\end{align}
where $t_{\rm m}$ (or redshift $z_{\rm m}$) is the lookback time corresponding to the merger of binary systems, $t_{\rm f}$ (or redshift $z_{\rm f}$) represents the time when binary systems form, and the time-delay distribution $P(t_{\rm d})$ represents the probability density of the binary system formed at time $t_{\rm f}$ and merged at time $t_{\rm m}$, with $t_{\rm d}=t_{\rm m}-t_{\rm f}$. The cosmic star formation rate (SFR) $R_{\rm sf}$ is assumed the Madau-Dickinson SFR \cite{Madau:2014bja}, with an exponential time delay \cite{Vitale:2018yhm}.

In the present work, we simulate the BNS and binary black hole (BBH) mergers for the following discussions. For BNS mergers, we consider the local comoving merger rate $R_{\rm m}(0)$ to be $R_{\rm m}(0)=320$ $\rm {Gpc^{-3}\ yr^{-1}}$, which is the estimated median rate based on the O3a observation\footnote{Note that the latest estimated median rate of BNS mergers by GWTC-3 is $105.5$ $\rm {Gpc^{-3}\ yr^{-1}}$ \cite{KAGRA:2021duu}. The rate adopted in this work falls within the range of the latest estimate, allowing the analysis to be considered as a specific scenario within that range.} \cite{LIGOScientific:2020ibl}. While for BBH mergers, we consider $R_{\rm m}(0)=23.9\ \rm {Gpc^{-3}\ yr^{-1}}$, which is the estimated median rate of O2 observation \cite{LIGOScientific:2020kqk} and is also consistent with the O3 observation \cite{KAGRA:2021duu}.

The sky location ($\theta$, $\phi$), the binary inclination angle $\iota$, the polarization angle $\psi$, and the coalescence phase $\psi_{\rm c}$ are randomly sampled in the ranges of $\cos\theta\in[-1,1]$, $\phi \in [0,2\pi]$, $\cos\iota \in[-1,1]$, $\psi \in[0,2\pi]$,
and $\psi_{\rm c} \in[0,2\pi]$, respectively. The redshift is sampled from the distribution that comes from the normalized observed merger rate. The coalescence time $t_{\rm c}$ is chosen as $t_{\rm c} \in [0,10]$ years for both bright and dark sirens. For the mass distribution, we employ the numerical fitting formulas to fit the curves shown in figure~7 of the POWER PEAK population model (mass of NS in the range of $[1.2, 2]~M_{\odot}$) in ref.~\cite{KAGRA:2021duu} for NS and the POWER LAW $+$ PEAK model (mass of BH in the range of $[5, 44]~M_{\odot}$) in refs.~\cite{LIGOScientific:2020kqk,KAGRA:2021duu} for BH.
{Note that in the present work, all the simulated BNS mergers will be used for the bright siren analysis, while all the BBH mergers will be used for the dark siren analysis.}

\subsection{Detections of GW events}\label{sec2.2}
{In this work, we consider the phenomenological non-spinning inspiral-merger-ringdown PhenmonA (IMRPhenmonA) model \cite{Cho:2015dra,Kumar:2016dhh}. The frequency-domain GW waveform is given by}
\begin{align}
\tilde{h}(f)={\mathcal{A}}\exp[i(2\pi ft_{\rm c}-\pi/4-2\psi_{\rm c}+2{\Psi}(f/2)-\varphi_{(2.0)})],
\label{eq:hf}
\end{align}
{where the amplitude $\mathcal{A}$ is given by}
\begin{equation}
{\mathcal{A}=\mathcal{A}_{\rm eff}\left\{
\begin{aligned}
&\left(\frac{f}{f_0}\right)^{-7/6}, &f < f_0, \\
&\left(\frac{f}{f_0}\right)^{-2/3}, &f_0 \leq f < f_1,\\
&\omega\mathcal{L}(f, f_1, f_2), &f_1 \leq f < f_3,
\end{aligned}
\right.}
\end{equation}
{and the form of $\mathcal{A}_{\rm eff}$ is written as}
\begin{align}
{\mathcal{A}_{\rm eff}}=
\frac{1}{d_{\rm L}}\sqrt{F_+^2\big(1+\cos^2(\iota)\big)^2+4F_\times^2\cos^2(\iota)}\times \sqrt{5\pi/96}\mathcal{M}_{\rm chirp}^{5/6}(\pi {f_0})^{-7/6},
\end{align}
$\Psi(f)$ and $\varphi_{(2,0)}$ are given by \cite{Sathyaprakash:2009xs,Blanchet:2004bb}
\begin{align}
&\Psi(f)=\frac{3}{256 \eta} \sum_{i=0}^{7} \alpha_{i}(2\pi Mf)^{(i-5) / 3}, \label{Psi}\\
&\varphi_{(2,0)}=\tan ^{-1}\left(-\frac{2 \cos (\iota) F_{\times}}{\big(1+\cos ^{2}(\iota)\big) F_{+}}\right),
\end{align}
where $d_{\rm L}$ is the luminosity distance to the GW source, $\iota$ is the inclination angle between the orbital angular momentum axis of the binary and the line of sight, $\mathcal{M}_{\rm chirp}$ is the observed chirp mass which is defined as $\mathcal{M}_{\rm chirp}=(1+z)M\eta^{3/5}$, $M=m_{1} + m_{2}$ is the total mass of the binary system, $\eta = m_1 m_2/(m_1+m_2)^2$ is the symmetric mass ratio, $t_{\rm c}$ is the coalescence time, $\psi_{\rm c}$ is the coalescence phase, and the coefficients $\alpha_{i}$ (we calculate the GW waveform to the 3.5 PN order) are detailedly given in e.g., refs.~\cite{Sathyaprakash:2009xs,Arun:2004hn}. 
{Note that $f_0$ is the merger frequency, $f_1$ is the ringdown frequency, $f_2$ is the decay-width of the ringdown, $f_3$ is the cut-off frequency, and the forms of $\omega$ and $\mathcal{L}$ are related to $f_0$, $f_1$, and $f_2$. The forms of $f_0$, $f_1$, $f_2$, $f_3$, $\omega$, and $\mathcal{L}$ are detailedly introduced in e.g., table~2 and eqs.~(21)--(23) of ref.~\cite{Robson:2018ifk}.}

Here $F_{+}$ and $F_{\times}$ are the antenna response functions of the detector, which are related to the location of the GW source ($\theta$, $\phi$) and the polarization angle $\psi$.
For Voyager and NEMO, the antenna response functions are
\begin{align}
F_+(\theta, \phi, \psi)=
\frac{1}{2}\big(1 + {\cos ^2}(\theta )\big)\cos (2\phi )\cos (2\psi ) 
- \cos (\theta )\sin (2\phi )\sin (2\psi ),\nonumber\\
F_\times(\theta, \phi, \psi)=
\frac{1}{2}\big(1 + {\cos ^2}(\theta )\big)\cos (2\phi )\sin (2\psi ) 
+ \cos (\theta )\sin (2\phi )\cos (2\psi ).
\label{eq:antenna}
\end{align}
When considering the detector network in the GW localization, it is optimal to consider the detailed detector coordinates in the antenna response functions. However, the coordinates of the 2.5G GW detectors are currently uncertain. Therefore, the above forms of $F_{+,\times}$ are adopted without considering the detector locations. In section~\ref{sec3.4}, we discuss the impact of the detector coordinates in the antenna response functions on the cosmological analysis.

\begin{figure}[!htbp]
\includegraphics[width=0.8\textwidth]{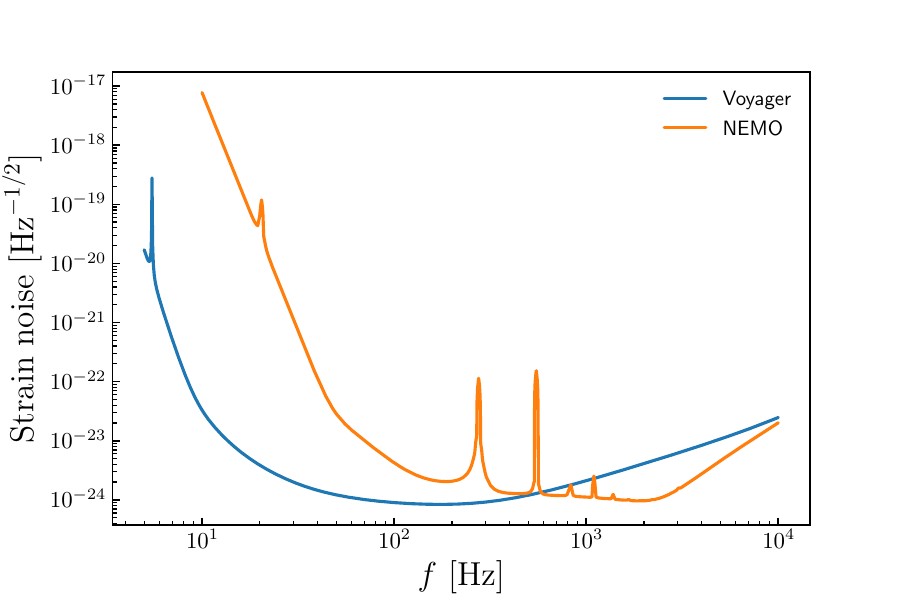}
\centering
\caption{Sensitivity curves of Voyager and NEMO adopted in this work.}\label{fig1}
\end{figure}

The signal-to-noise ratio (SNR) of the detector is given by
\begin{equation}
    \rho=(\tilde{h}|\tilde{h})^{1/2},
    \label{eq:rho}
\end{equation}
with the inner product being defined as
\begin{equation}
    (a|b)=2\int_{f_{\rm lower}}^{f_{\rm upper}}
    \frac{ a(f) b^{*}(f)+ a^{*}(f) b(f)}{S_{\rm n}(f)}\mathrm{d}f,
    \label{eq:innerproduct}
\end{equation}
where $f_{\rm lower}$ is the lower-frequency cutoff, $f_{\rm lower}$ is chosen as $f_{\rm lower}=\max(f_{\rm obs},5)$ Hz for NEMO and $f_{\rm lower}=\max(f_{\rm obs},10)$ Hz for Voyager, $f_{\rm obs}=(t_{\rm c}/5)^{-3/8}\mathcal{M}_{\rm chirp}^{-5/8}/8\pi$ is the observation frequency at the coalescence time $t_{\rm c}$, {$f_{\rm upper}=f_3$ is the upper-frequency cutoff,} $*$ denotes the complex conjugate, and $S_{\rm n}(f)$ is the one-side noise power spectral density (PSD). We adopt the forms of PSDs for NEMO from ref.~\cite{Ackley:2020atn} and Voyager from ref.~\cite{VOYAGER-SENSITIVITY}, as shown in figure~\ref{fig1}. Here the adopted Voyager sensitivity is BlueBird5 \cite{VOYAGER-bluebird5}, which lies between the latest two configurations of Voyager, i.e., Voyager deep and Voyager wideband \cite{VOYAGER-upgrade}.
The SNR threshold is set to 8 for bright sirens and 100 for dark sirens. 

We emphasize that the IMRPhenmonA GW waveform model we considered encodes only the primary and basic features of a GW signal. It lacks the typical tidal deformabilities present in the BNS mergers, the important effects of higher modes typically present in the BBH mergers, and spin parameters. Including these features in the GW waveform model would enable the possibility to break degeneracies between source parameters, thus improving measurement precisions of source parameters. A more robust analysis will be undertaken in future work.

\begin{figure}[!htbp]
\includegraphics[width=0.8\textwidth]{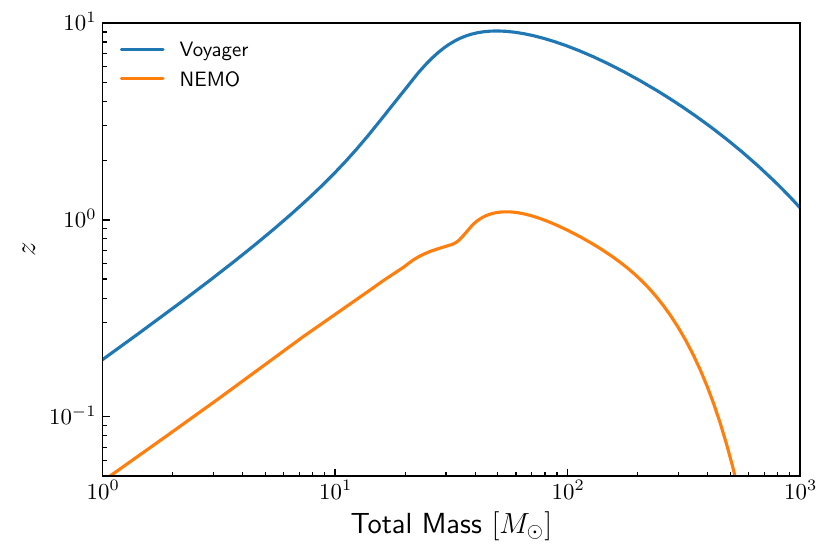}
\centering
\caption{{Detection horizons for the equal-mass non-spinning binaries as a function of the source frame total mass for Voyager and NEMO.}}\label{fig2}
\end{figure}

\begin{figure}[!htbp]
\includegraphics[width=0.8\textwidth]{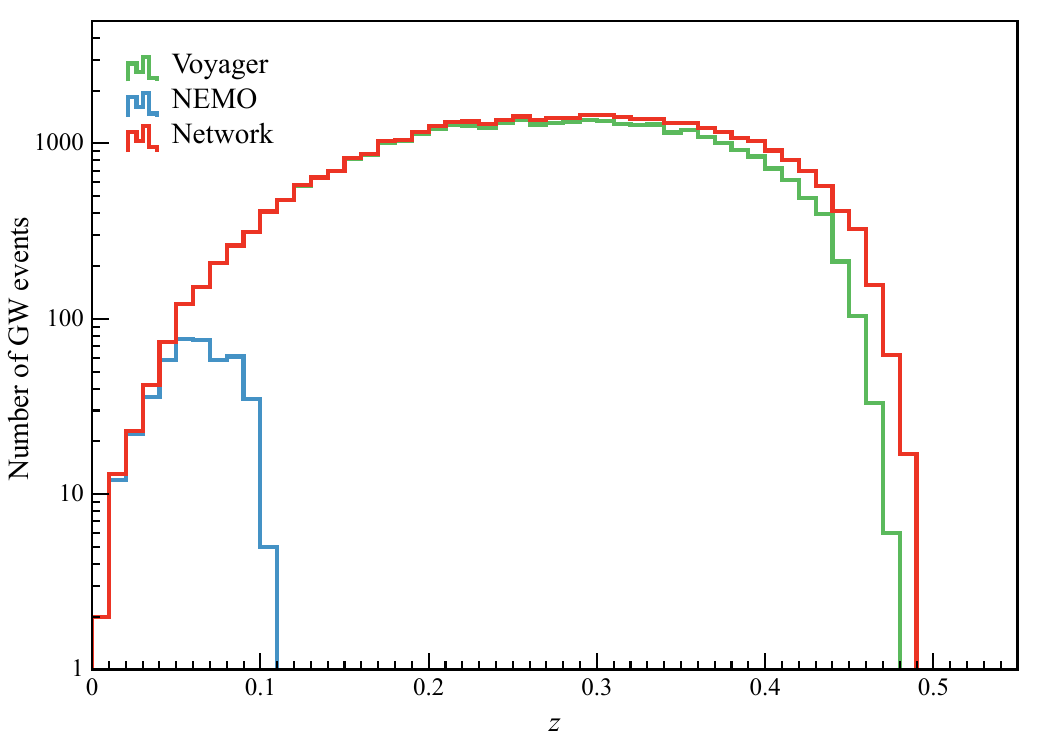}
\centering
\caption{{Redshift distributions of the BNS merger events detected by Voyager, NEMO, and the Voyager-NEMO network based on the 10-year observation.}}\label{fig3}
\end{figure}

{In figure~\ref{fig2}, we show the corresponding detection horizons ($\mathrm{SNR}\geq 8$) for equal-mass non-spinning binaries as a function of the source frame total mass for Voyager and NEMO. We can see that the detection capability of Voyager for compact binary coalescences is better than that of NEMO.}

{In figure~\ref{fig3}, we show the redshift distributions of the GW events detected by the Voyager-NEMO network, assuming the 10-year observation. We can see that the Voyager-NEMO network can detect BNS mergers at $z<0.5$. The detailed detection numbers are shown in table~\ref{tab1}.}

\subsection{Estimations of GW parameters}\label{sec2.3}

We use the Fisher information matrix to estimate the measurement errors of the GW parameters. For a set of parameters $\boldsymbol{\theta}$, the Fisher information matrix is given by
\begin{equation}
    F_{ij}=\left( \frac{\partial{\tilde {h}}}{\partial\theta_{i}}\bigg|\frac{\partial{\tilde{h}}}{\partial\theta_{j}}\right),
    \label{eq:fisher}
\end{equation}
where $\boldsymbol{\theta}$ denotes nine GW parameters ($d_{\rm L}$, $\mathcal{M}_{\rm chirp}$, $\eta$, $t_{\rm c}$, $\psi_{\rm c}$, $\iota$, $\theta$, $\phi$, $\psi$) for a binary system. The covariance matrix is equal to the inverse of the Fisher information matrix, i.e., $\mathrm{Cov}_{ij}=\sqrt{(F^{-1})_{ij}}$.
Then the measurement errors of the GW parameters is $\Delta\theta_{i}=\sqrt{(F^{-1})_{ii}}$.

In this work, considering only nine source parameters in the Fisher information matrix can reduce the computational time of source parameter estimations. 
In addition, due to the limitations of the Fisher information matrix, it performs rapidly and accurately in high-SNR cases but can be unstable in low-SNR cases \cite{Vallisneri:2007ev}. Given that the simulated SNR thresholds for BNSs and BBHs are 8 and 100, respectively, we check the instability of the Fisher matrix for BNS by printing the difference between $F\cdot F^{-1}$ and the identity matrix on the diagonal. This instability-check method is also adopted in \texttt{GWFAST} \cite{Iacovelli:2022mbg}.

The sky localization error is given by
\begin{align}
    \Delta{\Omega} = 2\pi |\sin\theta|\sqrt{(\Delta\theta)^2(\Delta\phi)^2-(\Delta\theta\Delta\phi)^2},
    \label{eq:location}
\end{align}
where $\Delta\theta$, $\Delta\phi$, and $\Delta\theta\Delta\phi$ can be calculated by the covariance matrix.

In addition to the measurement error caused by the GW observation, the errors from weak lensing and peculiar velocity are also considered. For the error caused by weak lensing, we adopt the form \cite{Tamanini:2016zlh,Speri:2020hwc,Hirata:2010ba}
\begin{align}
\sigma_{d_{\rm L}}^{\rm lens}(z)=\left[1-\frac{0.3}{\pi / 2} \arctan \left(z / 0.073\right)\right]
\times d_{\rm L}(z)\times 0.066\bigg[\frac{1-(1+z)^{-0.25}}{0.25}\bigg]^{1.8}.\label{lens}
\end{align}
The error caused by peculiar velocity is given by \cite{Kocsis:2005vv}
\begin{align}
\sigma_{d_{\mathrm{L}}}^{\rm{pv}}(z)=
d_{\rm L}(z) \times \left[  1+ \frac{c(1+z)^2}{H(z)d_{\rm L}}  \right]
\frac{	\sqrt{\left \langle v^2\right \rangle}}{c},
\label{eq:pvsigma}
\end{align}
where $H(z)$ is the Hubble parameter, $\sqrt{\left \langle v^2\right \rangle}$ is the peculiar velocity of the GW source, and $c$ is the speed of light in vacuum. Here we set $\sqrt{\left \langle v^2\right \rangle} = 500\ \rm{km}~\rm{s^{-1}}$, in agreement with the average value observed in galaxy catalogs \cite{He:2019dhl}.
Hence, the total error of $d_{\rm L}$ can be written as
\begin{equation}
\left(\sigma_{d_{\mathrm{L}}}\right)^{2}=\left(\sigma_{d_{\mathrm{L}}}^{\mathrm{inst}}\right)^{2}+\left(\sigma_{d_{\mathrm{L}}}^{\text {lens }}\right)^{2}+\left(\sigma_{d_{\mathrm{L}}}^{\rm {pv} } \right)^{2}.
\label{eq:totsigma}
\end{equation}

\subsection{Bright sirens: detections of SGRBs and inferences of cosmological parameters}\label{sec2.4}

In this subsection, we introduce the method of simulating the joint GW and short $\gamma$-ray burst (SGRB) events and performing cosmological analysis. We consider the THESEUS-like GRB detector in synergy with the 2.5G GW observation.

For the SGRB model, we consider the model of Gaussian-structure jet profile based on the GW170817/GRB170817A observation,
\begin{align}
    L_{\rm iso}(\theta_{\rm V})=L_{\rm on}\exp \left( -\frac{\theta_{\rm V}^2}{2\theta_{\rm c}^2}   \right),
    \label{eq:Gaussianjet}
\end{align}
where $L_{\rm iso}(\theta)$ is the isotropic equivalent luminosity of SGRB at the viewing angle $\theta_{\rm V}$, $L_{\rm on}$ is the on-axis isotropic luminosity defined by $L_{\rm on}=L_{\rm iso}(0)$, and $\theta_{\rm c}=4.7^\circ$ is the characteristic angle of the core. The direction of the jet is assumed to coincide with the direction of the angular momentum of the binary's orbital motion, i.e., $\iota=\theta_{\rm V}$.

The SGRB detectors are limited by the flux sensitivity for the distant GRB emissions at wider viewing angles. For the THESEUS-like detector, the sensitivity flux limit is $P_{\rm T}=0.2~\rm{ph~s^{-1}~cm^{-2}}$, in the 50-300 keV band. According to the relation between flux and luminosity for GRB~\cite{Meszaros:1995dj,Meszaros:2011zr}, we can convert the flux limit $P_{\rm T}$ to the luminosity by
\begin{equation}
L_{\rm iso}=4\pi d_{\rm L}^2(z)k(z)b/(1+z)P_{\rm T},
\end{equation}
where $k(z)$ is the $k$-correction form and is given by
\begin{align}
     k(z)=\int_{E{1}}^{E{2}}N(E)dE/ \int_{E1(1+z)}^{E2(1+z)} N(E)dE,\label{er:k_correction}
\end{align}
$b$ is an energy normalization and is given by
\begin{align}
    b=\int_{\rm 1~keV}^{\rm 10000~keV} EN(E)dE /\int_{E1}^{E2} N(E)dE,\label{eq:b_flux}
\end{align}
where $[E_1,E_2]$ is the detector's energy window, i.e., $E_1=50~{\rm keV}$ and $E_2=300~{\rm keV}$. The observed photon flux is scaled by $b$ to account for the missing fraction of the $\gamma$-ray energy seen in the detector band. The cosmological k-correction is due to the redshifted photon energy when traveling from source to detector. $N(E)$ is the observed GRB photon spectrum in units of $\rm{ph~s^{-1}~keV^{-1}~cm^{-2}}$.
For SGRBs, the function $N(E)$ is simulated by the Band function \cite{Band:2002te} which is a function of spectral indices $(\alpha_{\rm B}, \beta_{\rm B})$ and break energy $E_{\rm b}$,
\begin{equation}
\begin{array}{l}
N(E)=
\left\{\begin{array}{cc}
\begin{aligned}
    & N_{0} \left(\frac{E}{100~\mathrm{keV}}\right)^{\alpha_{\rm B} }\exp\left(-\frac{E}{E_{0}}\right), && E\leq E_{b}, \\
    &N_{0} \left(\frac{E_{b}}{100~\mathrm{keV}}\right)^{\alpha_{\rm B} - \beta_{\rm B}} \exp(\beta_{\rm B}-\alpha_{\rm B})\left(\frac{E}{100~\mathrm{keV}}\right)^{\beta_{\rm B}}, && E>E_{b},
    \end{aligned}
    \label{eq:NE}
\end{array}\right.
\end{array}
\end{equation}
where $E_{\rm b}=(\alpha_{\rm B}-\beta_{\rm B})E_0$ and $E_{\rm p}=(\alpha_{\rm B}+2)E_0$. Here we adopt $\alpha_{\rm B}=-0.5$, $\beta_{\rm B}=-2.25$, and a peak energy $E_{\rm p}=800~\rm keV$ in the source frame from ref.~\cite{Wanderman:2014eza}. This is a phenomenological fit to the observed spectra of GRB prompt emissions.

Finally, we could calculate the detection probability of SGRBs.
The detection probability of an SGRB is determined by $\Phi(L)dL$, where $\Phi(L)$ is the intrinsic luminosity function and $L$ is the peak luminosity of each burst, which is assumed to be isotropically emitted in the rest frame in the energy range of $1-10000~\rm keV$.
For the luminosity function, we assume a standard broken power law of the form when considering the SGRB
\begin{equation}
    \Phi(L)=
\left\{\begin{array}{cc}
\begin{aligned}
    &(\frac {L}{L^*})^{\alpha_{L}}, &&L<L^{*},\\
    &(\frac {L}{L^*})^{\beta_{L}}, &&L>L^{*},
    \end{aligned}
    \label{eq:PhiL}
\end{array}\right.
\end{equation}
where $L^{*}$ is the characteristic luminosity that separates the low and high ends of the luminosity function. Following ref.~\cite{Wanderman:2014eza}, we adopt the characteristic slopes $\alpha_{L}=-1.95$, $\beta_{L}=-3$ and $L^{*}=2\times10^{52}\ {\rm erg\ s^{-1}}$ to describe the characteristic luminosity for different regimes. Finally, for each BNS sample, we could select the SGRB detections from the BNS samples by calculating the probability of the detection $P$
\begin{equation}
    P=\int_{L_{\rm c}}^{\infty}\Phi(L){\rm d}L.
\end{equation}
We emphasize that we make an optimistic assumption that all the detectable GW-SGRB events can provide redshift measurements. {According to the calculations, the numbers of the simulated bright sirens from Voyager, NEMO, and the Voyager-NEMO network is 42, 5, and 50, respectively, based on the 10-year observation, as shown in table~\ref{tab1}. The numbers of GW-SGRB events from Voyager and NEMO are also consistent with those given in refs.~\cite{Chen:2020zoq,Sarin:2021qqo}. Similar discussions but for the third-generation (3G) GW detectors are also discussed in e.g., refs.~\cite{Ronchini:2022gwk,Han:2023exn}. For example, compared to the results reported in ref.~\cite{Han:2023exn}, we see that the numbers of detected GW events from the 3G GW detectors are two orders of magnitudes greater than the 2.5G result, and the number of the 3G GW-SGRB events exceeds the 2.5G results by one order of magnitude.}

\begin{table}
\caption{\label{tab1} {Estimated numbers of the GW events detected by Voyager, NEMO, and the Voyager-NEMO network, and the joint GW-SGRB events detected by THESEUS-like GRB detector, assuming the 10-year observation.}}
\centering
\renewcommand{\arraystretch}{1.5}
\begin{tabular}{c|c|c}
\hline\hline
Detection strategy	& GW events	&GW-SGRB events\\ \hline   
{Voyager}&{35715} &{42}\\
{NEMO}&{442} &{5}\\
{Network}&   {39249}    &       {50}       \\ 
    \hline\hline
\end{tabular}
\end{table}

{For the cosmological parameter estimations using the mock bright siren data, we adopt the Markov-chain Monte Carlo analysis to maximize the likelihood $\mathcal{L}\propto(-\chi^2/2)$ and infer the posterior probability distributions of cosmological parameters $\vec{\Omega}$. The $\chi^2$ function is defined as} 
\begin{equation}
{\chi^2=\sum_{i=1}^{N}\left(\frac{\hat{d}_{\mathrm{L},i}-d_{\rm L}(z_i,\vec{\Omega})}{\sigma_{d_{{\rm L},i}}}\right)^2,}
\end{equation}
{where $N$ is the number of the bright siren data. We use the {\tt emcee} Python module \cite{Foreman-Mackey:2012any} to constrain the cosmological parameters.} 

\subsection{Galaxy localization}\label{sec2.5}
We first obtain the sky localization errors and luminosity distance errors of BBHs. For the simulation of galaxy catalogs, we uniformly simulate in a co-moving volume with a number density of $0.02~\mathrm{Mpc}^{-3}$ in the range of $z\in [0,0.2]$ (note that the adopted galaxy number is in the middle of the observational error bars, see figure~1
of ref.~\cite{Barausse:2012fy} for more details).
By combining the flat priors for $\Omega_{\rm m}\in [0.1, 0.5]$ and $H_0\in [60, 80]$~km s$^{-1}$ Mpc$^{-1}$, and the luminosity distance errors, we obtain the lower and upper limits of redshift ($z_{\rm min}$, $z_{\rm max}$), assuming the $\Lambda$CDM model. We then derive the $2\times 2$ covariance matrix $\mathrm{Cov}$ including $\theta$ and $\phi$ from the whole covariance matrix and use the new covariance matrix to calculate $\chi^2$, which describes the angular deviation from an arbitrary galaxy to the GW source and is given by
\begin{equation}
\chi^{2}=\boldsymbol{\xi}\left(\operatorname{Cov}^{-1}\right) \boldsymbol{\xi}^{\top}=\sum_{i, j} \xi_{i}\left(\operatorname{Cov}^{-1}\right)_{i j} \xi_{j},\label{chi2}
\end{equation}
with $\boldsymbol{\xi}=(\theta-\bar{\theta},\phi-\bar{\phi})$. Here ($\bar{\theta}$, $\bar{\phi}$) represent the angular location of the GW source, and ($\theta$, $\phi$) represent the angular location of an arbitrary galaxy. We adopt the boundary of the GW source's angular localization with $\chi^2=9.21$, corresponding to the 99\% confidence level. Therefore, if the galaxies satisfy $\chi^2\leq 9.21$ and $z\in [z_{\rm min}, z_{\rm max}]$, they can be considered as the potential host galaxies of the GW source.

\subsection{Dark sirens: Bayesian inferences of cosmological parameters}\label{sec2.6}

\begin{figure}[!htbp]
\includegraphics[width=0.8\textwidth]{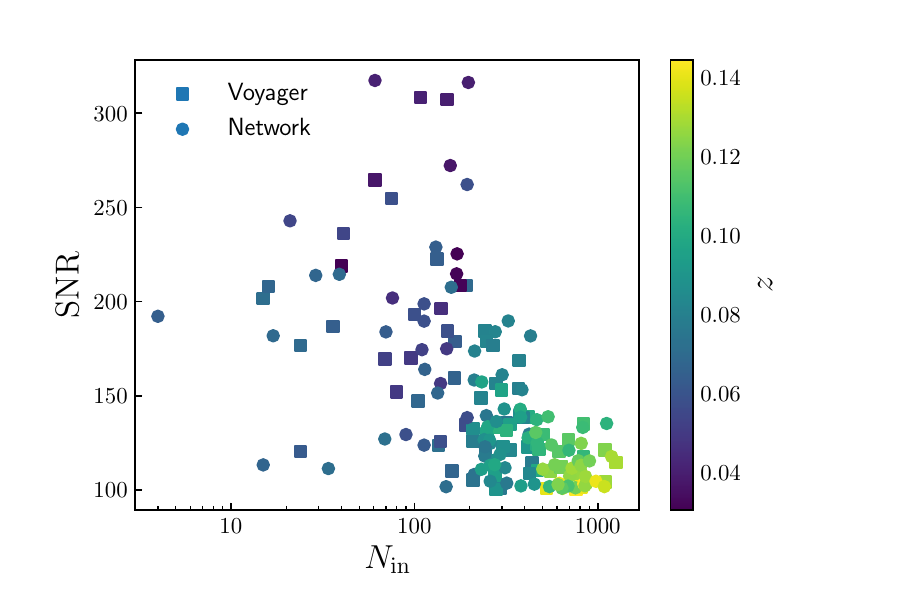}
\centering
\caption{Correlation between SNRs of dark sirens from Voyager and the Voyager-NEMO network, the number of the potential host galaxies $N_{\rm in}$, and redshifts of dark sirens based on the 10-year observation.}\label{fig4}
\end{figure}

For the GW events that cannot identify EM counterparts to determine the redshift information, we need to use the cross-matching of GW localization and galaxy catalogs to obtain redshift information. In this subsection, we briefly introduce the method of using dark sirens to infer cosmological parameters.
Using the method introduced in section~\ref{sec2.5}, we calculate the number of the potential host galaxies $N_{\rm in}$ and show the SNR--$N_{\rm in}$ plot in figure~\ref{fig4}. We can see that the higher the SNR of the GW event, the smaller the number of potential host galaxies $N_{\rm in}$. Meanwhile, the GW events with lower SNRs mainly have higher redshifts.

We use the following Bayesian method and the {\tt emcee} Python module \cite{Foreman-Mackey:2012any} to constrain the cosmological parameters.
Here we consider the future galaxy survey projects, e.g., the China Space Station Telescope, which are anticipated to yield a galaxy catalog with a completeness level of 100\% at redshifts $z\leq 0.3$ \cite{Song:2022siz}. Hence, we do not consider the galaxy catalog's incompleteness in the following methods, as we only use GW events with $z<0.15$. The posterior probability distribution of the cosmological parameters $\vec{\Omega}$ is
\begin{equation}
  p(\vec{\Omega}|\vec{X}_{\rm GW})\propto p(\vec{\Omega})\mathcal{L}(\vec{X}_{\rm GW}|\vec{\Omega}),
\end{equation}
where $p(\vec{\Omega})$ are the prior distributions of the cosmological parameters and $\vec{X}_{\rm GW}$ represent the GW data.
Assuming that the GW events are independent of each other, the likelihood of the GW data can be expressed as
\begin{equation}
    \mathcal{L}(\vec{X}_{\rm GW}|\vec{\Omega})=\prod_{i=1}^{N_{\rm GW}}\mathcal{L}(X_{{\rm GW},i}|\vec{\Omega}),
\end{equation}
where $N_{\rm GW}$ is the number of GW events. The likelihood function $\mathcal{L}(X_{{\rm GW}}|\vec{\Omega})$ is obtained by marginalizing the redshifts of the GW events,
\begin{equation}
    \mathcal{L}(X_{\rm GW}|\vec{\Omega})=\frac{1}{\beta(\vec{\Omega})}\int p(X_{\rm GW}|d_{\rm L}(z,\vec{\Omega}))p_{z}(z){\rm d}z ,\label{eq:likelihood}
\end{equation}
where $p(X_{\rm GW}|d_{\rm L}(z,\vec{\Omega}))$ is the posterior distribution of GW event's $d_{\rm L}$ obtained from the GW observation
and is given by
\begin{equation}
\label{con:GW likelihood}
    \begin{aligned}
    p\left( X_{\rm GW}|d_{\rm L}(z,\vec{\Omega})\right)=\frac{1}{\sqrt{2\pi}\sigma _{d_{\rm L}}}\exp\left[-\frac{(\hat{d}_{\rm L}-d_{\rm L}(z,\vec{\Omega}))^2}{2\sigma _{d_{\rm L}}^2}\right ],
    \end{aligned}
\end{equation}
where $\hat{d}_{\rm L}$ is the observed luminosity distance of the GW source and the luminosity distance error $\sigma_{d_{\rm L}}$ can be obtained from eq.~(\ref{eq:totsigma}).
$p_{z}(z)$ are the prior distributions of the redshifts of GW sources and can be expressed as,
\begin{equation}
   p_{z}(z)\propto \frac{1}{N_{\rm in}}\sum_{n=1}^{N_{\rm in}}W_{n}\delta(\hat{z}_n-z),
\end{equation}
where $\delta(\hat{z}_n-z)$ is the delta function with $\hat{z}_n$ being the redshift of the $n$-th potential host galaxy and $W_{n}$ is the angular position weight of the $n$-th potential host galaxy
\begin{equation}
    W_n\propto \exp\left(-\frac{1}{2}\chi^2_n\right).
\end{equation}
Here, $\chi^2_n$ is obtained from eq.~(\ref{chi2}). The term $\beta(\vec{\Omega})$ in the denominator of eq.~(\ref{eq:likelihood}) represents the correction for the selection effect arising from GW observations, as given by
\begin{equation}
    \beta(\vec{\Omega})=\int p_{\rm{det}}^{\rm{GW}}\left(d_{\rm{L}}(z, \vec{\Omega})\right) p_z(z) \mathrm{d} z,
\end{equation}
where $p_{\rm{det}}^{\rm{GW}}\left(d_{\rm{L}}(z, \vec{\Omega})\right)$ represents the probability of detecting a GW event at $d_{\rm{L}}(z, \vec{\Omega})$, which is obtained by marginalizing over the GW source parameters while fixing $d_{\rm L}$ \cite{Gray:2019ksv}.

We emphasize that the application of dark siren in constraining cosmological parameters heavily relies on the localization ability, i.e., the number of potential host galaxies, as introduced in section~\ref{sec2.5}. Therefore, we adopt the SNR threshold of dark sirens to be 100, which can ensure more precise localizations of dark sirens, thereby significantly reducing the number of potential host galaxies. According to the calculations of the 10-year observation, the detection numbers of BBH merger events are 83 and 93 for Voyager and the Voyager-NEMO network, respectively. Owing to the fact that NEMO is not sensitive to the GW signals from BBH mergers, we do not consider the single NEMO observatory in the dark siren analysis. 

In addition, since the peculiar velocity of galaxies is considered in eq.~(\ref{eq:pvsigma}), the calculation of $p_{z}(z)$ results in a delta function. If the peculiar velocity in $d_{\rm L}$ is not considered, it would instead be a Gaussian function. Note that clustering is not considered. In reality, GW events are likely to merge in crowded environments where the local galaxy number density deviates from the global average. This deviation leads, on average, to a larger number of galaxies per error volume, which would in turn affect the likelihood for the cosmological parameters and change the final constraint results of cosmological parameters. Since we uniformly simulate the galaxy catalogs in a co-moving volume, this clustering effect cannot be considered. Consequently, the dark siren results in this work are optimistic. We will consider this effect in future work.


\begin{table*}[!htbp]
\renewcommand\arraystretch{1.5}
\caption{The absolute errors (1$\sigma$) and the relative errors of the cosmological parameters in the $\Lambda$CDM, $w$CDM, and $w_{0}w_{a}$CDM models {using the Voyager, NEMO, the Voyager-NEMO network, CMB, CMB+Voyager, CMB+NEMO, and CMB+Network data. Note that the mock GW data are based on the 10-year observation.} Here $H_0$ is in units of km s$^{-1}$ Mpc$^{-1}$.}\label{tab2}
\centering
\normalsize
\setlength{\tabcolsep}{6mm}{
\resizebox{\textwidth}{!}{
\begin{tabular}
{p{0.9cm}<{\centering} p{0.4cm}<{\centering} p{0.7cm}<{\centering}p{0.7cm}<{\centering} p{0.7cm}<{\centering}p{0.6cm}<{\centering}p{2cm}<{\centering} p{1.8cm}<{\centering} p{2cm}<{\centering} }
\hline\hline
{Model} & {Error}  & {Voyager}&  {NEMO}&{Network} &  {CMB}     &{CMB+Voyager} &  {CMB+NEMO} &{CMB+Network} \\\hline
 \multirow{4}{*}{$\Lambda$CDM}&  ${\sigma(\Omega_{\rm m})}$&  {$-$}&{$-$}  &  {$-$}& {0.0085}     & {0.0079}& {0.0084}     & {0.0073} \\
& ${\sigma(H_0)}$     & {1.40} & {4.40} & {1.10} & {0.61}   & {0.57}& {0.61}  & {0.52} \\
& ${\varepsilon(\Omega_{\rm m})}$  & {$-$}  & {$-$}& {$-$} & {$2.69\%$}     &  {$2.49\%$} & {$2.65\%$}&  {$2.30\%$} \\
& ${\varepsilon(H_0)}$   & {2.08\%}& {6.44\%} & {1.63\%} & {$0.90\%$}    & {0.85\%} & {$0.90\%$}& {0.77\%}\\
\hline
\multirow{6}{*}{$w$CDM}&  ${\sigma(\Omega_{\rm m})}$ & {0.260} & {$-$}  & {0.250} & {0.070} & {0.015}& {0.034}  & {0.011}\\
& ${\sigma(H_0)}$  & {2.25} & {4.15}  & {1.75} & {7.75}     & {1.50} & {3.50}       & {1.10}\\
& ${\sigma(w)}$   & {0.670}& {$-$}  & {0.585} & {0.260}     & {0.055}& {0.120}      & {0.045}\\
& ${\varepsilon(\Omega_{\rm m})}$& {$50.98\%$} & {$-$}&  {$54.35\%$}& {$21.52\%$}    &{$4.72\%$}& {$10.47\%$} & {$3.47\%$} \\
& ${\varepsilon(H_0)}$    & {3.40\%}& {6.17\%}  & {2.63\%} & {$11.40\%$}   & {2.23\%}  & {$5.19\%$}    & {1.63\%}\\
& ${\varepsilon(w)}$   & {$72.83\%$}& {$-$}& {$56.80\%$}& {$25.74\%$}    & {$5.53\%$}& {$12.05\%$}  & {$4.50\%$} \\
\hline
\multirow{7}{*}{$w_{0}w_{a}$CDM} &  ${\sigma(\Omega_{\rm m})}$  & {0.290} & {$-$}& {0.265} & {0.079}& {0.020} & {0.037} & {0.016}\\
& ${\sigma(H_0)}$  &{2.60} &{4.40} &{1.95}&{9.00}   & {2.00} &{4.05}   & {1.60}\\
& ${\sigma(w_{0})}$   & {1.03} & {$-$} & {0.83} & {0.59}     & {0.46} & {0.53}       & {0.44}\\
& ${\sigma(w_{a})}$  & {$-$} & {$-$}& {$-$} & {$-$}  & {1.70}& {$-$} & {1.60} \\
& ${\varepsilon(\Omega_{\rm m})}$ & {$55.77\%$}& {$-$} & {$51.96\%$}  & {$24.84\%$}  & {$6.06\%$}& {$11.56\%$} & {$4.89\%$}\\
& ${\varepsilon(H_0)}$   & {3.90\%} & {6.48\%} & {2.91\%}& {$13.04\%$}      & {3.03\%}& {$6.01\%$}  & {2.41\%} \\
& ${\varepsilon(w_{0})}$  & {78.03\%} & {$-$}  & {60.14\%}  & {$94.40\%$}     & {88.46\%}& {$94.64\%$} & {67.69\%} \\
\hline\hline
\end{tabular}}}
\end{table*}

\section{Results and discussions}\label{sec3}

In this section, we shall report the constraint results of the cosmological parameters. To ensure a comprehensive and robust analysis of cosmological parameter estimations, we consider two scenarios, bright sirens (optimistic) and dark sirens (conservative). For the optimistic scenario, we use the simulated bright siren data to constrain the $\Lambda$CDM, $w$CDM, and $w_0w_a$CDM models. {Furthermore, we consider three detection strategies, Voyager, NEMO, and the Voyager-NEMO network to make the following discussions. In order to show the ability of the bright sirens to break the cosmological parameter degeneracies generated by CMB, we also show the constraint results of CMB and the combination of GW and CMB.}
For the CMB data, we employ the ``Planck distance priors" from the Planck 2018 observation \cite{Planck:2018vyg,Chen:2018dbv}. The constraint results for the cosmological parameters of interest are shown in figures~\ref{fig5}--\ref{fig8} and summarized in table~\ref{tab2}. We use $\sigma(\xi)$ and $\varepsilon(\xi)$ to represent the absolute and relative errors of parameter $\xi$, with $\varepsilon(\xi)$ defined as $\varepsilon(\xi)=\sigma(\xi)/\xi$. For the conservative scenario, we use the mock dark sirens from Voyager and the Voyager-NEMO network to constrain the $\Lambda$CDM model. The constraint results are shown in figure~\ref{fig9}, and summarized in table~\ref{tab3}. {We discuss the impact of uncertainty in the BNS merger rate on estimating cosmological parameters. The related constraint results are shown in figures~\ref{fig10} and \ref{fig11}, and summarized in table~\ref{tab4}.} Furthermore, we also discuss the impact of the detector coordinates in the antenna response functions on estimating $H_0$ in the dark siren scenario. The distributions of $\sigma_{d_{\rm L}}/d_{\rm L}$ and $\Delta\Omega$ are shown in figure~\ref{fig12} and the constraint results are shown in figure~\ref{fig13} and summarized in table~\ref{tab5}.

\subsection{Bright sirens}\label{sec3.1}

\begin{figure}[!htbp]
\includegraphics[width=0.6\textwidth]{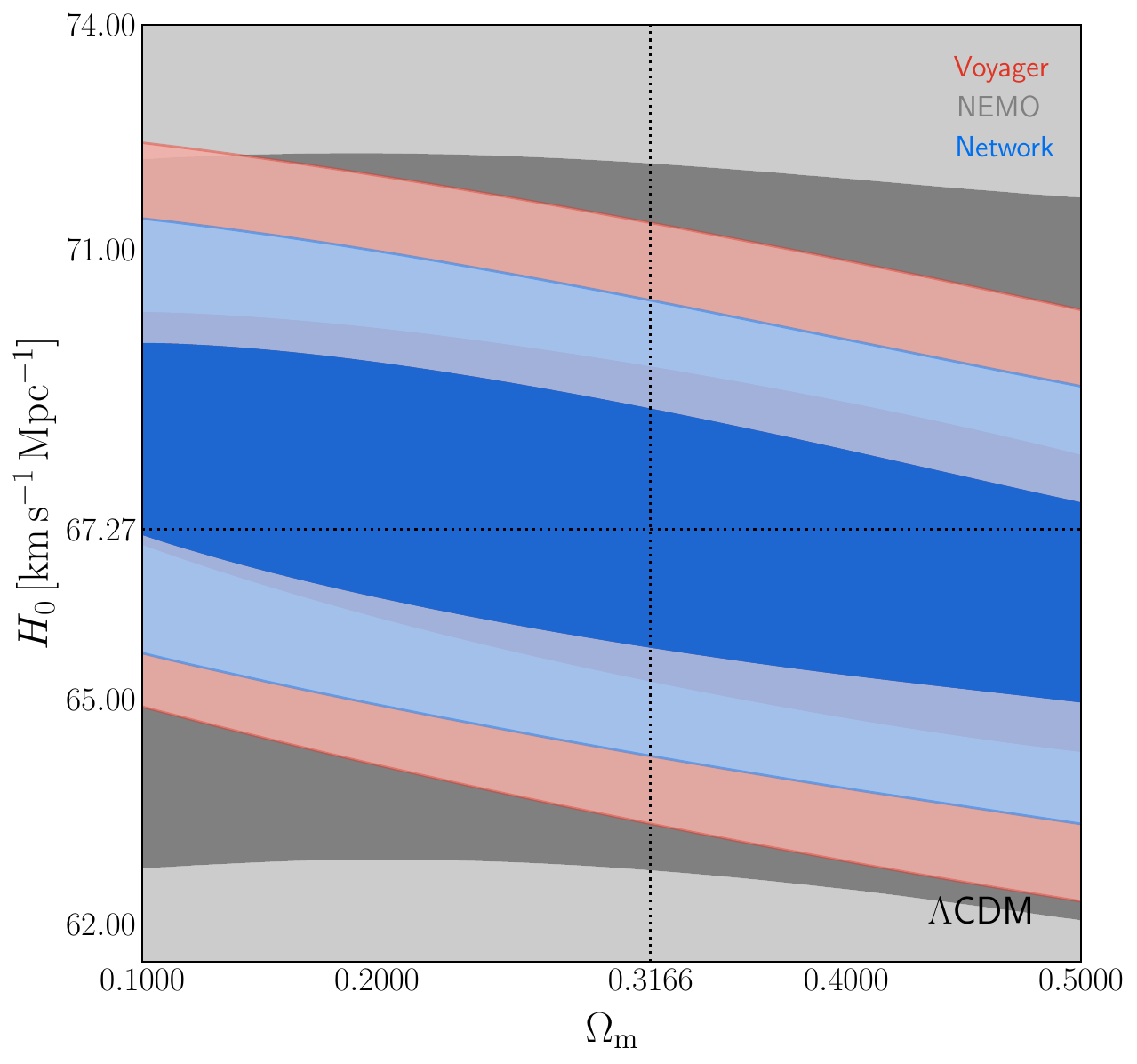}
\centering
\caption{{Two-dimensional marginalized contours ($68.3\%$ and $95.4\%$ confidence level) in the $\Omega_{\rm m}$--$H_0$ plane using the 10-year mock bright siren data from Voyager, NEMO, and the Voyager-NEMO network in the $\Lambda$CDM model. Here, the dotted lines indicate the fiducial values of cosmological parameters preset in the simulation.}}\label{fig5}
\end{figure}

\begin{figure}[!htbp]
\includegraphics[width=0.6\textwidth]{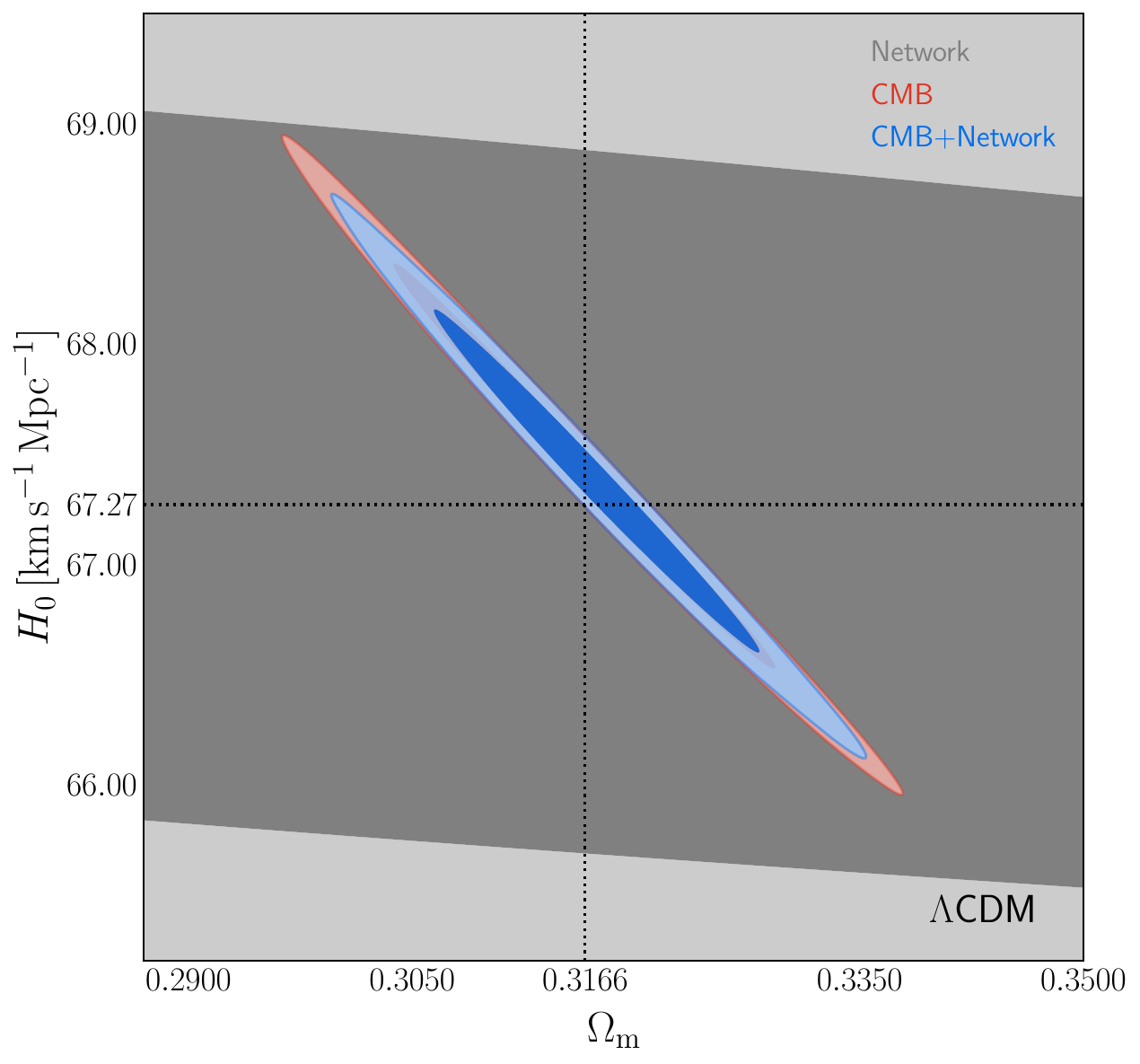}
\centering
\caption{{Same as figure~\ref{fig5}, but using the Network, CMB, and CMB+Network data.}}\label{fig6}
\end{figure}

\begin{figure*}[!htbp]
\includegraphics[width=0.48\textwidth]{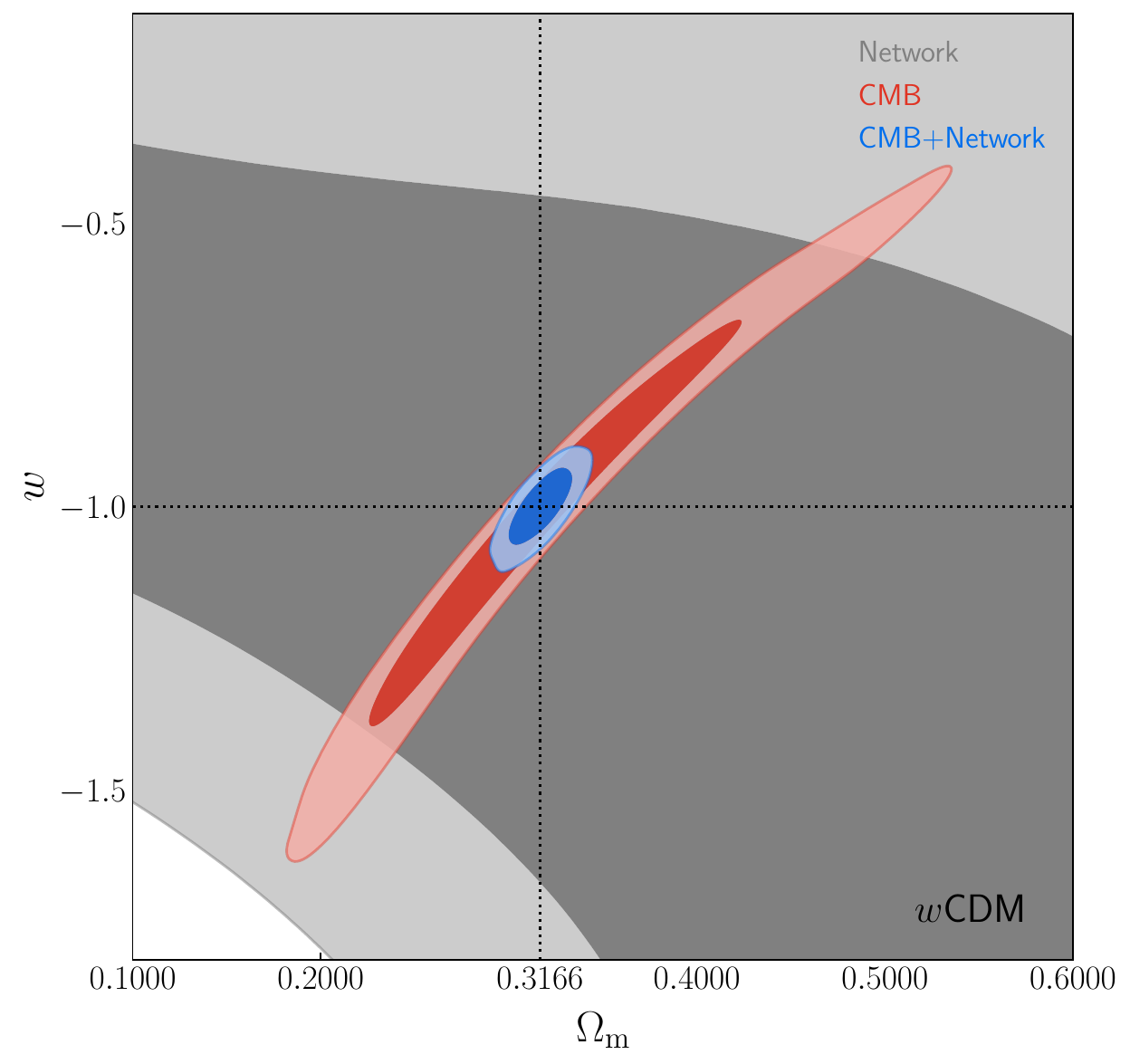}
\includegraphics[width=0.48\textwidth]{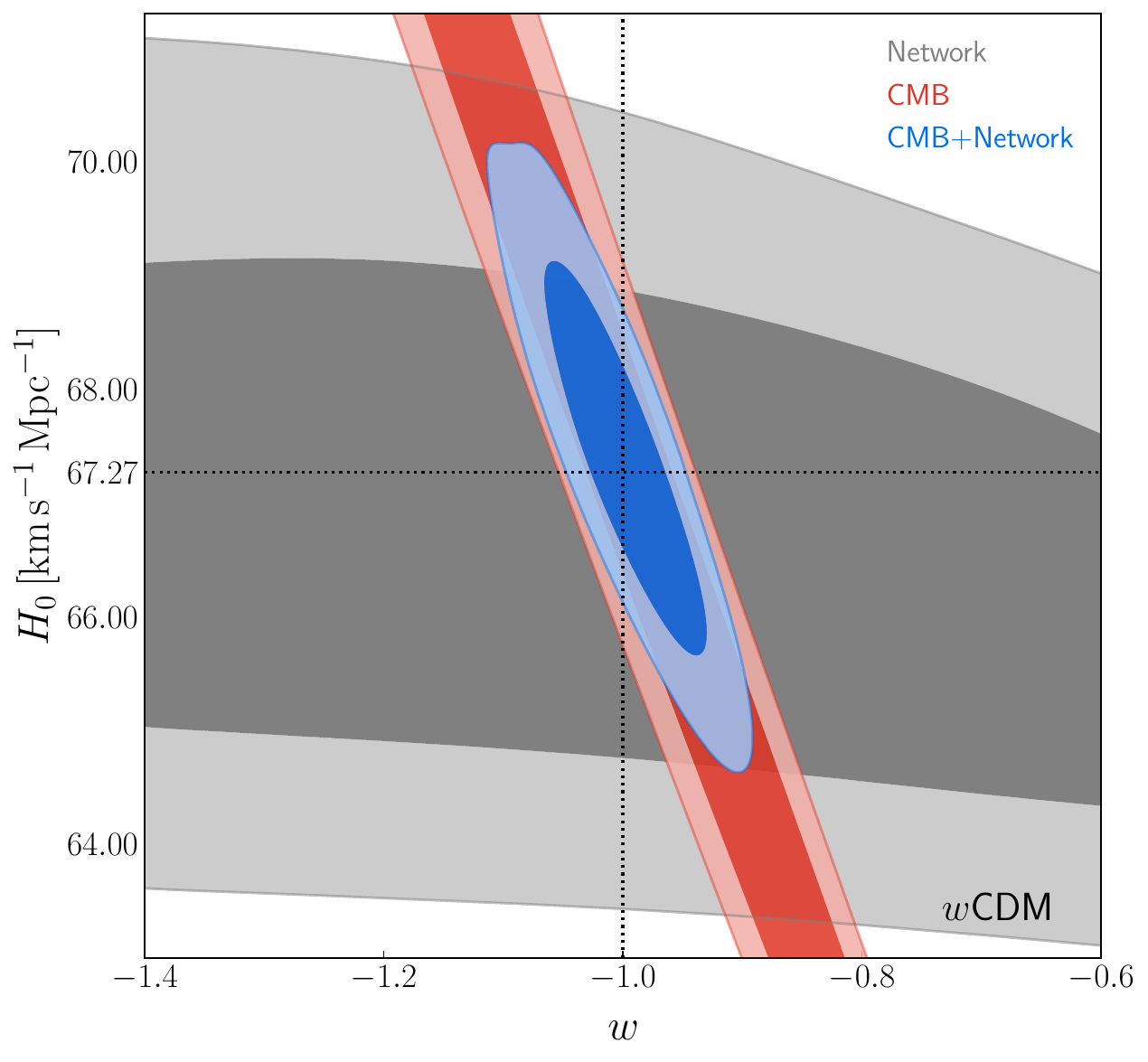}
\centering
\caption{{Same as figure~\ref{fig5}, but for the $w$CDM model in the $\Omega_{\rm m}$--$w$ and $w$--$H_0$ planes.}}\label{fig7}
\end{figure*}

\begin{figure}[!htbp]
\includegraphics[width=0.6\textwidth]{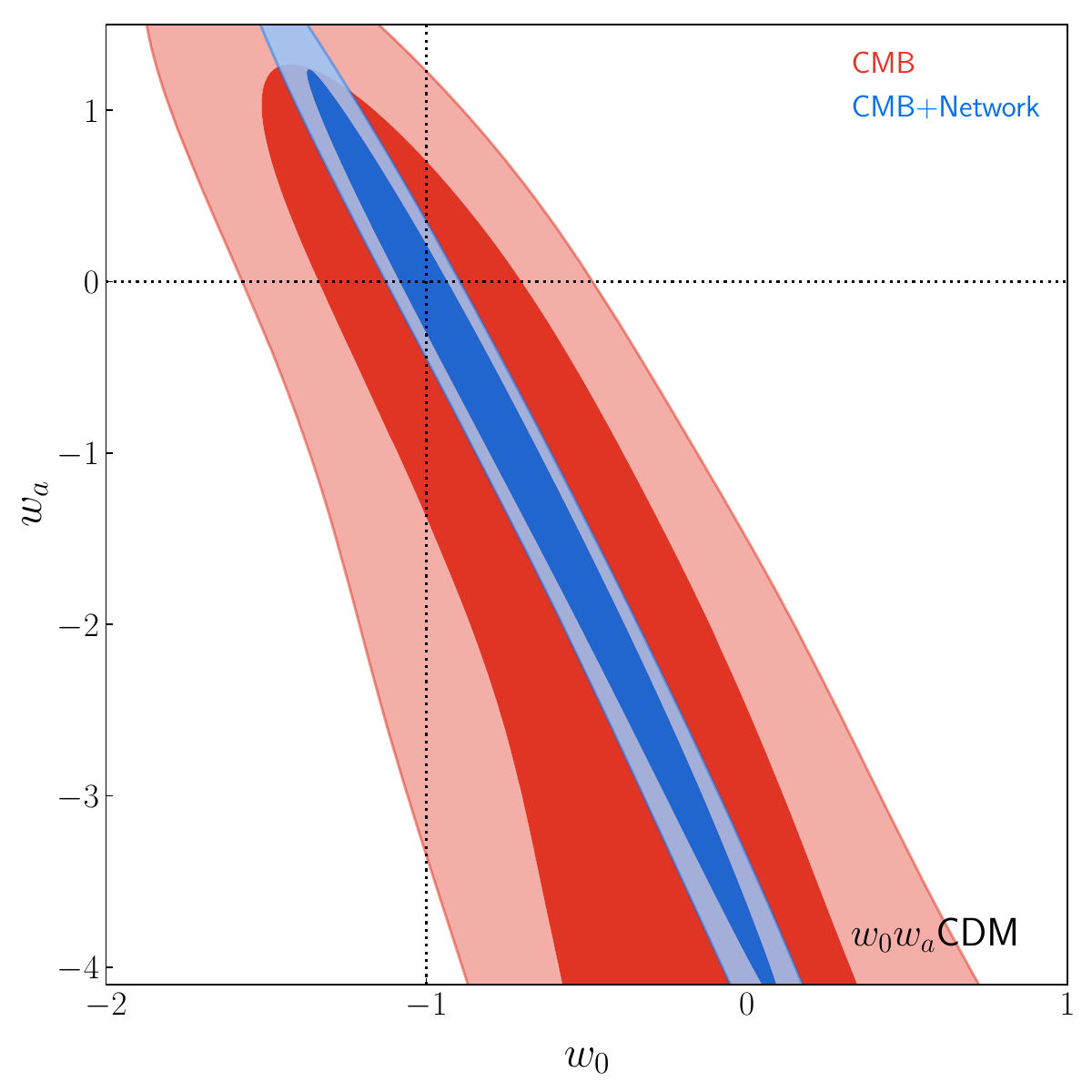}
\centering
\caption{{Same as figure~\ref{fig5}, but using the CMB and CMB+Network data for the $w_0w_a$CDM model in the $w_0$--$w_a$ plane.}}\label{fig8}
\end{figure}

{In figure~\ref{fig5}, we show the constraint results from Voyager, NEMO, and the Voyager-NEMO network. We can see that Network gives the best constraint results, followed by Voyager and NEMO. Since the GW-SGRB event number of Voyager is about 8 times more than that of NEMO, Voyager gives much better constraints than those of NEMO. Concretely, Voyager and NEMO give $\sigma(H_0)=1.40$ km s$^{-1}$ Mpc$^{-1}$ and $\sigma(H_0)=4.40$ km s$^{-1}$ Mpc$^{-1}$ with precisions of 2.08\% and 6.44\%, respectively. When considering the detector network, the constraint precision of $H_0$ is 21.4\% and 75\% better than those from the single Voyager and NEMO, respectively. Furthermore, when considering the Voyager-NEMO network, the constraint precision of $H_0$ is improved to 1.63\%, which is close to the distance ladder measurement \cite{Riess:2021jrx}. Therefore, in the 2.5G GW detector era, it is more meaningful to consider the detector network, i.e., the Voyager-NEMO network to perform cosmological analysis. In addition, the constraint precision of $H_0$ given by the 2.5G detector network is much better than the current constraint result given by the LIGO-Virgo-KAGRA observations \cite{LIGOScientific:2017adf,LIGOScientific:2021aug}. However, the constraint results are much worse than the forecast results by the 3G GW detectors (measurement precisions in the order of better than 1\%) in e.g., refs.~\cite{Belgacem:2019tbw,Zhang:2019loq,Jin:2020hmc,Han:2023exn}. Note that in the following we compare our bright siren results with those reported in ref.~\cite{Han:2023exn}, which adopts the same bright siren analysis method as the present work. Since the Voyager-NEMO network gives the best constraints on cosmological parameters, we take the Voyager-NEMO network as the representative in the following discussions.}

{In figure~\ref{fig6}, we show the constraint results in the $\Lambda$CDM model using the Network, CMB, and CMB+Network data. As clearly seen, bright sirens from the Voyager-NEMO network can give tight constraint on $H_0$, with a precision of 1.63\%. 
Since GW could break the cosmological parameter degeneracies generated by CMB, the combination of them could improve the measurement precisions of cosmological parameters. The combination of CMB and Network gives $\sigma(\Omega_{\rm m})=0.0073$ and $\sigma(H_0)=0.52$ km s$^{-1}$ Mpc$^{-1}$, which are 14.1\% and 14.8\% better than those of CMB, respectively. Furthermore, the constraint precisions of $\Omega_{\rm m}$ and $H_0$ using the CMB+Network data are 2.30\% and 0.77\%, respectively, which are worse than the results by the 3G GW detector network given in ref.~\cite{Han:2023exn}.}

{In figure~\ref{fig7}, we show the constraint results in the $\Omega_{\rm m}$--$w$ and $w$--$H_0$ planes for the $w$CDM model. We see that it is challenging to constrain the $w$CDM model by solely using the 2.5G GW bright sirens. Although the 3G GW detectors cannot precisely measure $w$ as well, even the worst constraint from the Einstein Telescope (ET) is $\sigma(w)=0.138$ \cite{Han:2023exn}, which is 76.4\% better than that of Voyager-NEMO. In addition, the contours of CMB and Network show different parameter degeneracy orientations (almost orthognal in the $w$--$H_0$ plane), and thus the combination of them could effectively break the cosmological parameter degeneracies. The combination of CMB and Network gives $\sigma(\Omega_{\rm m})=0.011$, $\sigma(H_0)=1.10$ km s$^{-1}$ Mpc$^{-1}$, and $\sigma(w)=0.045$, which are 84.3\%, 85.8\%, and 82.7\% better than those of CMB, respectively. For the EoS parameter of dark energy, the precision can reach 4.50\% using the combination of CMB and 2.5G GW bright sirens.}

{Since the constraint results by solely using the GW bright sirens are rather poor, we only show the combination results in the $w_0w_a$CDM model. In figure~\ref{fig8}, we show the case in the $w_0$--$w_a$ plane for the $w_0w_a$CDM model using the CMB and CMB+Network data. Similarly, the constraints on $w_0$ and $w_a$ by the 2.5G GW detectors are also much worse than those of the 3G GW detectors. The constraint on $w_0$ by ET is 74.8\% better than that of the Voyager-NEMO network \cite{Han:2023exn}.
We see that the combination of CMB and 2.5G GW bright sirens can also break cosmological parameter degeneracies. The joint CMB+Network data give $\sigma(w_0)=0.44$, which is 25.4\% better than that of CMB. Furthermore, the single GW or CMB data cannot constrain $w_a$, the combination of CMB and Network gives $\sigma(w_a)=1.6$, worse than the result by the Cosmic Explorer (CE) \cite{Han:2023exn}.}

{The above results show that the bright sirens from the 2.5G GW detectors can potentially help resolve the Hubble tension, but it is more helpful to consider the 2.5G GW detector network to perform the cosmological analysis. Nevertheless, it is challenging to precisely measure other cosmological parameters by solely using the 2.5G GW bright sirens. Fortunately, the parameter degeneracy orientations of CMB and bright sirens are different, and thus the combination of them could break the cosmological parameter degeneracies, especially in the dynamical dark energy models. In addition, the ability of the Voyager-NEMO network to constrain cosmological parameters is much worse than that of the 3G GW detectors. The constraint precisions on $H_0$ using the 3G GW detectors can improve by an order of magnitude compared to that obtained from the 2.5G GW detectors.}

\subsection{Dark sirens}\label{sec3.2}


\begin{figure*}[!htbp]
\includegraphics[width=0.96\textwidth]{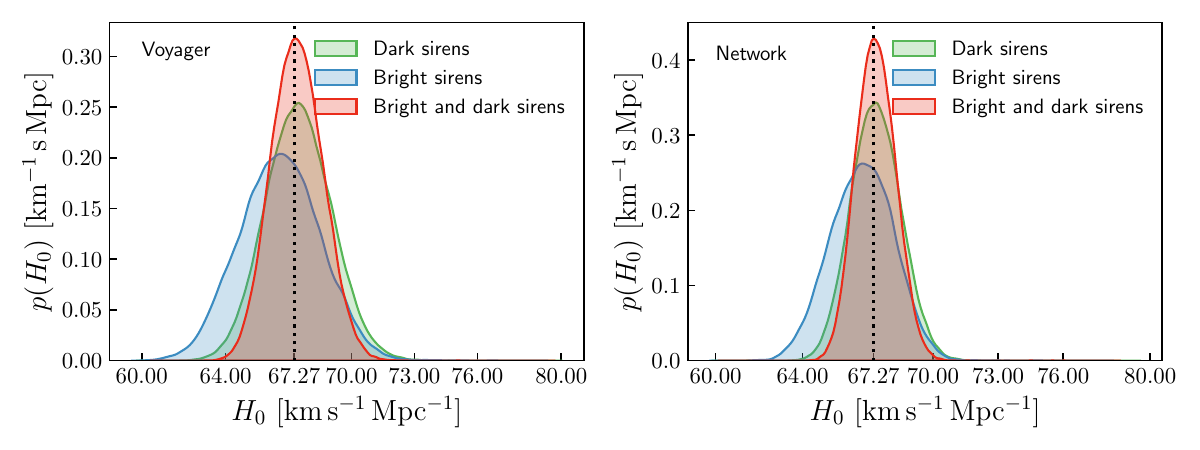}
\centering
\caption{One-dimensional marginalized probability distributions for $H_0$ from dark sirens, bright sirens, and bright and dark sirens based on the 10-year observation of Voyager and the Voyager-NEMO network. Here the galaxy catalogs are uniformly simulated in a co-moving volume with a number density of $0.02~\mathrm{Mpc^{-3}}$, as described in section~\ref{sec2.5}. The dotted line indicates the fiducial values of cosmological parameters preset in the simulation.}\label{fig9}
\end{figure*}

\begin{table*}
\caption{{Constraint results of the dark sirens and the combination of bright and dark sirens from Voyager and the Voyager-NEMO network based on the 10-year observation. The first column represents the detection strategy, the second column represents the number of the simulated dark siren data used in this work, the third to fifth columns represent the percentages of GW events with $N_{\text {in}} \leq 10$, $N_{\text {in}} \leq 100$, and $N_{\text {in}} \leq 1000$, the sixth and seventh column represents the relative errors of $H_0$ using dark sirens and the combination of bright and dark sirens.} See section~\ref{sec3.1} for more details about bright sirens. {Note that the relative errors of $\Omega_{\rm m}$ are not shown due to the poor constraints.}}
\centering
\normalsize
\renewcommand{\arraystretch}{1.5}
\setlength{\tabcolsep}{4mm}{
\resizebox{\textwidth}{!}{
\begin{tabular}
{p{3cm}<{\centering} p{1.5cm}<{\centering} p{1.5cm}<{\centering} p{1.5cm}<{\centering}p{2cm}<{\centering} p{3cm}<{\centering} p{4cm}<{\centering}}
\hline\hline
\multirow{2}{*}{{Detection strategy}}&\multirow{2}{*}{{Number}}&	\multirow{2}{*}{{$N_{\text {in }} \leq 10$}}&\multirow{2}{*}{{$N_{\text {in}} \leq 100$}}&	\multirow{2}{*}{{$N_{\text {in }} \leq 1000$}}&\multicolumn{2}{c}{{$\varepsilon(H_0)$}}\\\cline{6-7} 
&&&&&{Dark sirens}&{Bright and dark sirens}\\
\hline
{Voyager}&{83}&{1.20\%}&{18.07\%}&{96.39\%}&2.32\%&1.83\%\\
{Network}&{93}&{1.08\%}&{13.98\%}&{96.77\%}&1.70\%&1.35\%\\
\hline \hline \label{tab3}
\end{tabular}}}
\end{table*}

In this subsection, we shall report the constraint results of the dark sirens. In figure~\ref{fig9}, we show the constraint results of the scenarios of dark sirens, bright sirens, and bright and dark sirens from Voyager and the Voyager-NEMO network. We can see that Voyager could measure $H_0$ to a precision of 2.32\%. 
The dark sirens from the Voyager-NEMO network can potentially constrain $H_0$ to a precision of 1.70\%. Such precision is also one order of magnitude worse than those from the 3G GW detectors. In addition, $\Omega_{\rm m}$ cannot be well constrained due to the low-redshift dark sirens ($z<0.15$). 



In addition, LIGO-Virgo-KAGRA also gives the constraint on $H_0$ using the combination of bright and dark sirens with a precision of about 10\% \cite{LIGOScientific:2021aug}. Therefore, we also forecast the constraint on $H_0$ using the bright and dark sirens from the 2.5G GW detectors. The bright and dark sirens from the single Voyager observatory give a 1.83\% measurement on $H_0$.
The constraint precision of $H_0$ is expected to reach 1.35\% using the bright and dark sirens from the Voyager-NEMO network, better than the result of the distance ladder measurement \cite{Riess:2021jrx}. We see that compared to the single Voyager, {the constraint precision of $H_0$ using the Voyager-NEMO network can improve from 2.32\% to 1.70\%, as the addition of NEMO enhances the localization ability. For the combined bright and dark siren scenario, the improvement is from 1.83\% to 1.35\%.}
In conclusion, it is worth expecting that the standard siren observations from the 2.5G GW detectors can play a crucial role in making an arbitration for the Hubble tension.

\subsection{{Impact of uncertainty in the BNS merger rate on estimating cosmological parameters}}\label{sec3.3}

\begin{figure*}[!htbp]
\includegraphics[width=0.48\textwidth]{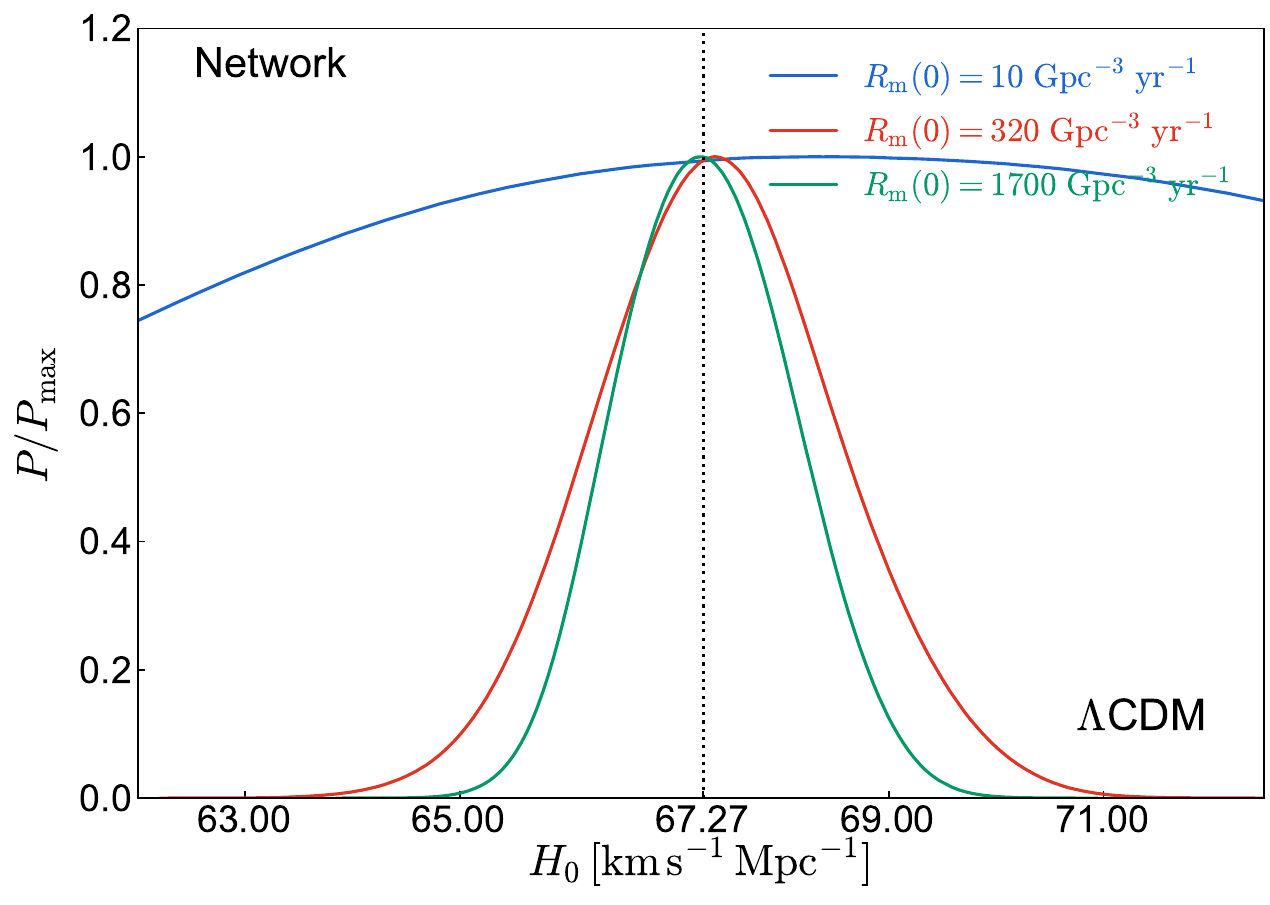}
\includegraphics[width=0.48\textwidth]{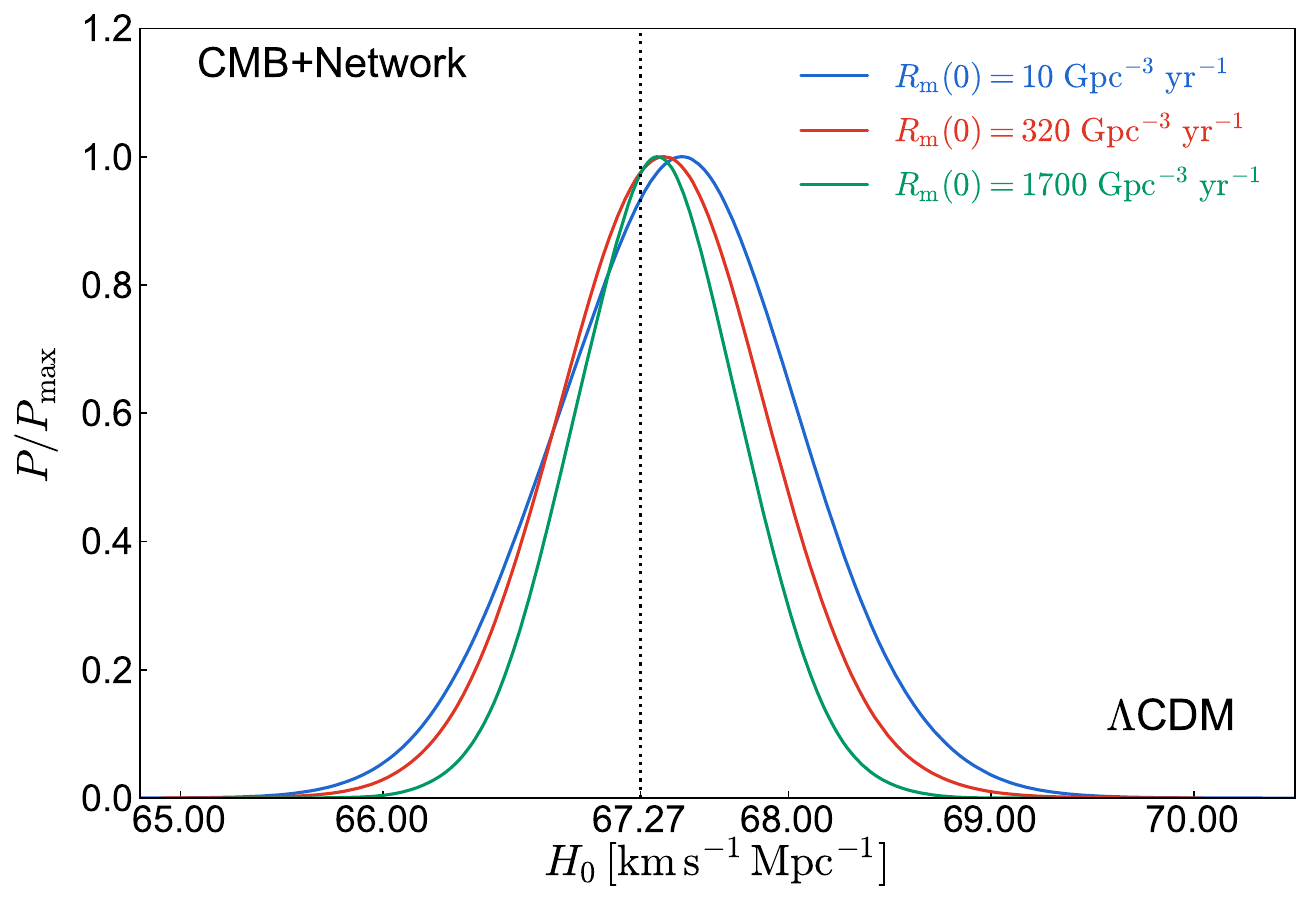}
\centering
\caption{{One-dimensional marginalized probability distributions for $H_0$ in the $\Lambda$CDM model based on the Voyager-NEMO network (10-year observation) and CMB+Network data in the scenarios of considering the BNS merger rates $R_{\rm m}(0)=10$ $\rm {Gpc^{-3}\ yr^{-1}}$, $R_{\rm m}(0)=320$ $\rm {Gpc^{-3}\ yr^{-1}}$, and $R_{\rm m}(0)=1700$ $\rm {Gpc^{-3}\ yr^{-1}}$. Here, the dotted lines indicate the fiducial values of cosmological parameters preset in the simulation.}}\label{fig10}
\end{figure*}

\begin{figure*}[!htbp]
\includegraphics[width=0.48\textwidth]{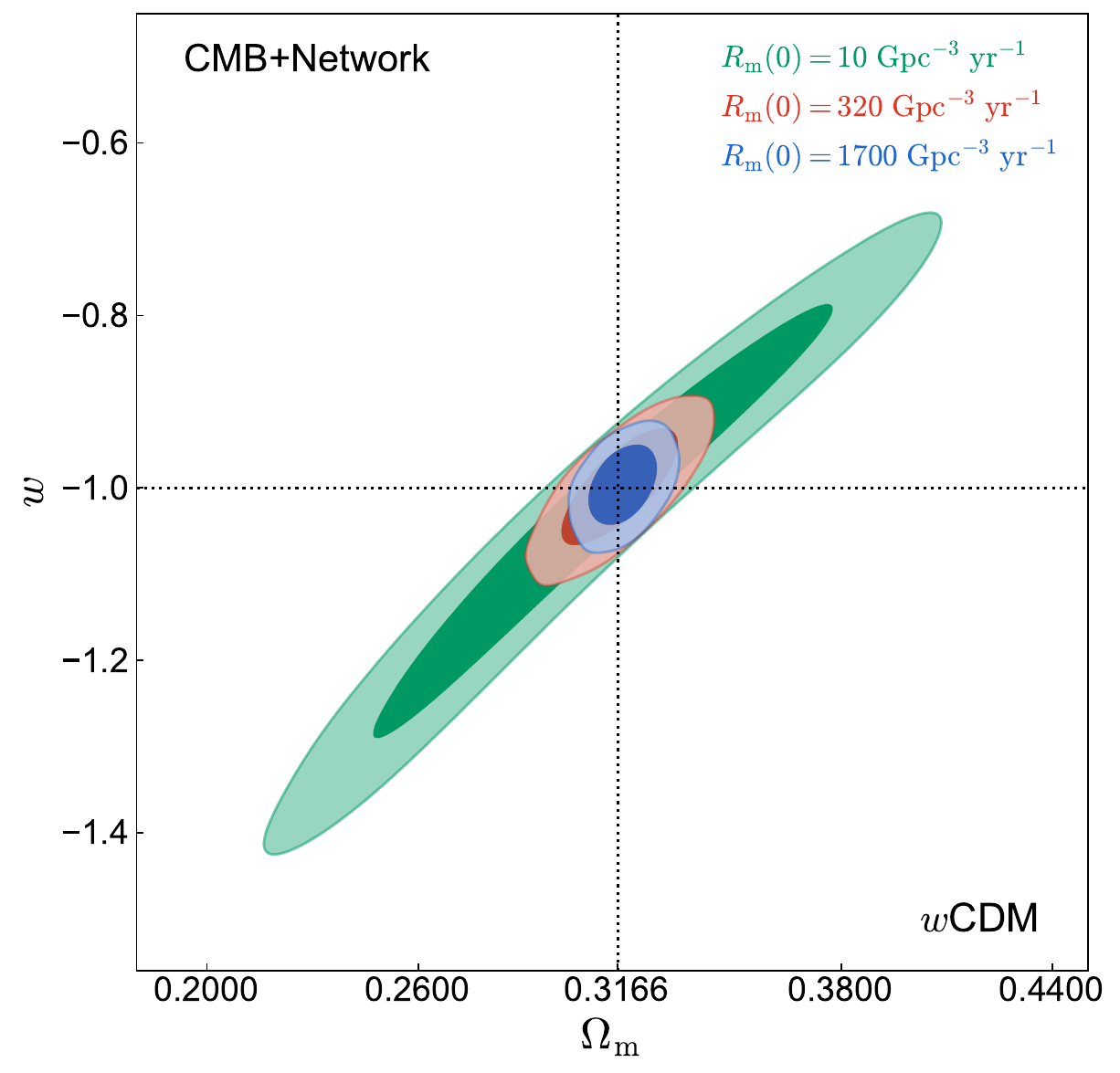}
\includegraphics[width=0.48\textwidth]{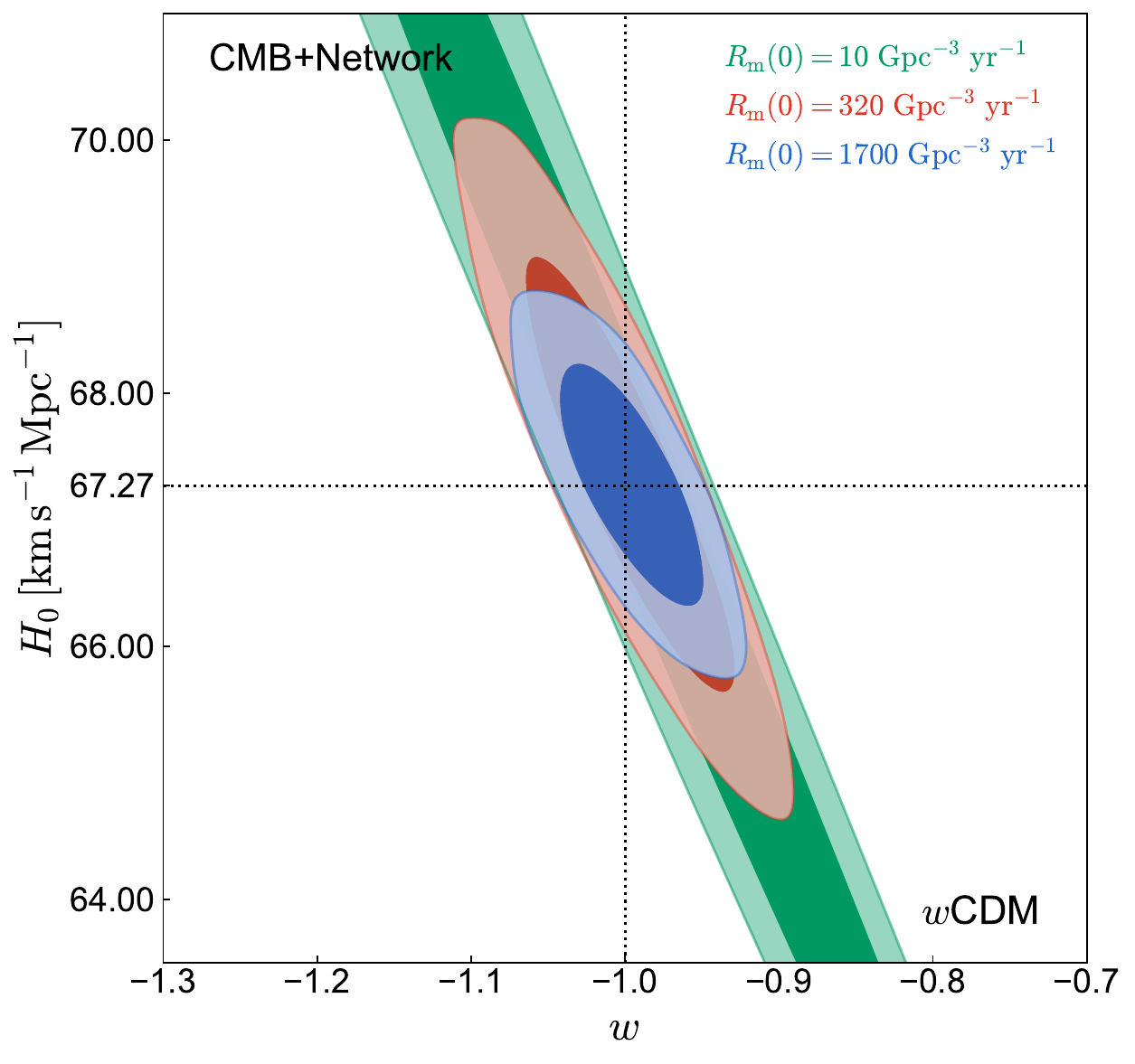}
\centering
\caption{{Two-dimensional marginalized contours ($68.3\%$ and $95.4\%$ confidence level) in the $\Omega_{\rm m}$--$w$ and $w$--$H_0$ planes in the $w$CDM model from the CMB+Network data in the scenarios of considering the BNS merger rates $R_{\rm m}(0)=10$ $\rm {Gpc^{-3}\ yr^{-1}}$, $R_{\rm m}(0)=320$ $\rm {Gpc^{-3}\ yr^{-1}}$, and $R_{\rm m}(0)=1700$ $\rm {Gpc^{-3}\ yr^{-1}}$. Here, the dotted lines indicate the fiducial values of cosmological parameters preset in the simulation.}}\label{fig11}
\end{figure*}

\begin{table*}
\caption{\label{tab4} {Same as table~\ref{tab2}, but using the Network and CMB+Network data based on the BNS merger rates $R_{\rm m}(0)=10\ \rm {Gpc^{-3}\ yr^{-1}}$ and $R_{\rm m}(0)=1700\ \rm {Gpc^{-3}\ yr^{-1}}$. Here $H_0$ is in units of km s$^{-1}$ Mpc$^{-1}$. Note that constraints on cosmological parameters based on $R_{\rm m}(0)=320\ \rm {Gpc^{-3}\ yr^{-1}}$ are shown in table~\ref{tab2}.}}
\centering
\normalsize
\renewcommand{\arraystretch}{1.5}
\setlength{\tabcolsep}{4mm}{
\resizebox{\textwidth}{!}{
\begin{tabular}
{p{1.1cm}<{\centering} p{1.0cm}<{\centering} p{2.5cm}<{\centering} p{2.5cm}<{\centering}p{0.2cm}<{\centering} p{2.5cm}<{\centering} p{2.5cm}<{\centering}}
\hline\hline
\multirow{2}{*}{{Model}}&	\multirow{2}{*}{{Error}}& \multicolumn{2}{c}{{Network}}& &\multicolumn{2}{c}{{CMB+Network}}\\ \cline{3-4}\cline{6-7} 
&  &{$R_{\rm m}(0)=10$}&{$R_{\rm m}(0)=1700$}& &{$R_{\rm m}(0)=10$}	&{$R_{\rm m}(0)=1700$}\\ \hline

\multirow{4}{*}{{$\Lambda$CDM}}&{$\sigma(\Omega_{\rm m})$}&{$-$}&{0.0990}& &{0.0085}&{0.0059}\\
& {$\sigma(H_0)$}&{5.30}&{0.82}& &{0.61}&{0.42}\\
& {$\varepsilon(\Omega_{\rm m})$}&{$-$}&{31.94\%}& &{2.69\%}&{1.86\%}\\
& {$\varepsilon(H_0)$}  &{7.59\%}&{1.21\%} & &{0.90\%}&{0.62\%}\\ \hline

\multirow{6}{*}{{$w$CDM}}&{$\sigma(\Omega_{\rm m})$}&{$-$}&{0.1950}& &{0.0410}&{0.0064}\\
& {$\sigma(H_0)$}&{$-$}&{1.10}& &{4.85}&{0.64}\\
& {$\sigma(w)$}& {$-$}&{0.460}& &{0.160} &{0.031}\\
& {$\varepsilon(\Omega_{\rm m})$}&{$-$}&{51.32\%}& &{13.18\%}&{2.01\%}\\
& {$\varepsilon(H_0)$}  &{$-$}&{1.64\%}& &{7.09\%}&{0.95\%}\\ 
&{$\varepsilon(w)$}& {$-$}&{40.35\%}& &{15.53\%} &{3.11\%}\\
    \hline\hline
\end{tabular}}}
\end{table*}

{We investigate the impact of uncertainty in the BNS merger rate on the cosmological parameter estimations. Since the current inferred BBH merger rate has a small error range (no difference in magnitude), only BNS mergers are considered in this subsection. We consider the lower, medium, and upper limits of BNS merger rates given by the latest O3 observation from LIGO-Virgo-KAGRA \cite{KAGRA:2021duu}, i.e., $R_{\rm m}(0)=10$ $\rm {Gpc^{-3}\ yr^{-1}}$ (lower), $R_{\rm m}(0)=320$ $\rm {Gpc^{-3}\ yr^{-1}}$ (medium, as discussed above), and $R_{\rm m}(0)=1700$ $\rm {Gpc^{-3}\ yr^{-1}}$ (upper). A higher BNS merger rate leads to a higher detection rate of bright sirens and thus better constraints on the cosmological parameters.}

{In figure~\ref{fig10}, we show the constraint results using the Network and CMB+Network data in the $\Lambda$CDM model. From the left panel, as clearly seen, bright sirens based on the 
lower merger rate cannot constrain $H_0$. However, bright sirens based on the medium and upper merger rates can both give tight constraints on $H_0$. With upper merger rate, bright sirens from Network are expected to constrain $H_0$ to a precision of 1.21\%, better than the current distance ladder measurement \cite{Riess:2021jrx}. When combining with CMB, tight constraint on $H_0$ could also be obtained, but the combined results based on the lower merger rate are mainly contributions from CMB. With the addition of the Network data with upper merger rate to CMB, the constraints on $\Omega_{\rm m}$ and $H_0$ could be improved by 30.6\% and 31.1\%, respectively, because the parameter degeneracies generated by CMB are broken by GW.}

{Since the above results show that it is challenging to measure cosmological parameters in the dynamical dark energy models by solely using the bright sirens, we only show the case of the combination of GW and CMB, shown in figure~\ref{fig11}. The detailed constraint results are shown in table~\ref{tab4}. We can see that the ability of GW to break the cosmological parameter degeneracies is strong, even the bright sirens with lower merger rate can improve the cosmological parameter precisions to some extent. When combining the mock bright sirens from Network (lower merger rate) with CMB, the constraints on $\Omega_{\rm m}$, $H_0$, and $w$ can be improved by 41.4\%, 37.4\%, and 38.5\%, respectively. While for the upper merger rate case, the improvements on $\Omega_{\rm m}$, $H_0$, and $w$ are 90.9\%, 91.7\%, and 88.1\%, respectively. Moreover, the constraint precision of $w$ using the CMB+Network data with upper merger rate is 3.11\%, close to the latest CMB+SN result \cite{Brout:2022vxf}. We can conclude that the BNS merger rates can seriously affect the cosmological parameter estimations, but the 2.5G GW standard sirens hold promise in helping address the Hubble tension.}

\subsection{{Impact of the detector coordinates in the antenna response functions on estimating the Hubble constant in the dark siren scenario}}\label{sec3.4}
\begin{figure*}[!htbp]
\includegraphics[width=0.96\textwidth]{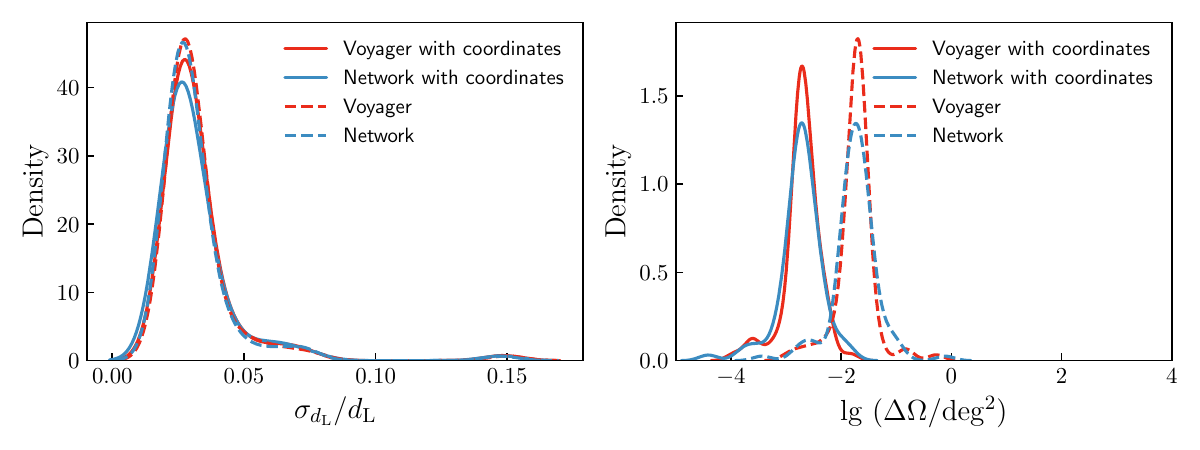}
\centering
\caption{{Distributions of $\sigma_{d_{\rm L}}/d_{\rm L}$ and $\Delta\Omega$ for simulated BBHs at $z<0.15$ for Voyager and the Voyager-NEMO network with and without considering the detector coordinates in the antenna response functions.}}\label{fig12}
\end{figure*}

\begin{figure*}[!htbp]
\includegraphics[width=0.96\textwidth]{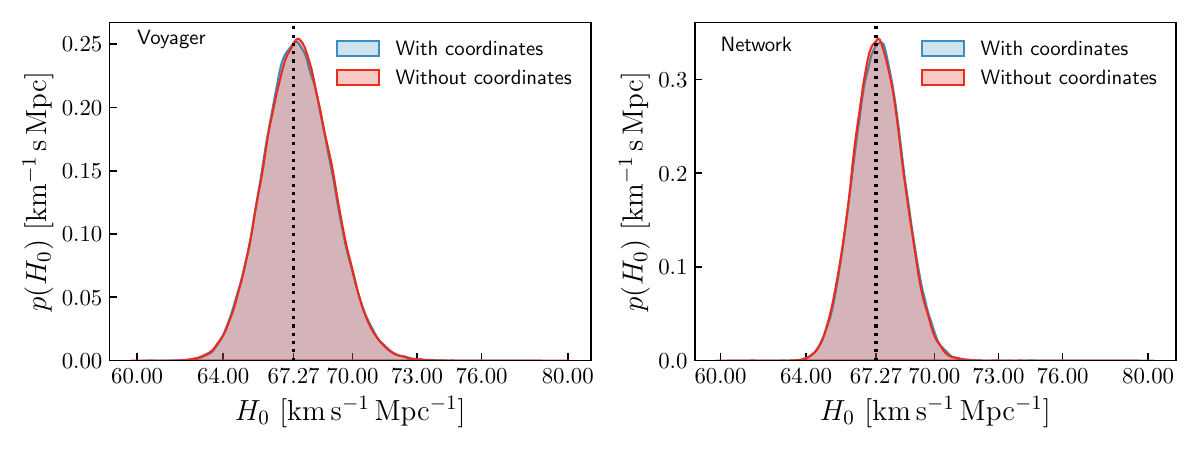}
\centering
\caption{{Same as figure~\ref{fig9}, but for the scenario with and without considering the detector coordinates in the antenna response functions.}}\label{fig13}
\end{figure*}

\begin{table*}[!htbp]
\caption{{Same as table~\ref{tab3}, but considers the detector coordinates in the antenna response functions for the dark siren results.}}
\centering
\normalsize
\renewcommand{\arraystretch}{1.5}
\setlength{\tabcolsep}{4mm}{
\resizebox{\textwidth}{!}{
\begin{tabular}
{p{4cm}<{\centering} p{1.5cm}<{\centering} p{2cm}<{\centering} p{2cm}<{\centering}p{2cm}<{\centering} p{1.5cm}<{\centering}}
\hline\hline
{{Detection strategy}}&{{Number}}&	{{$N_{\text {in }} \leq 10$}}&{{$N_{\text {in}} \leq 100$}}&	{{$N_{\text {in }} \leq 1000$}}&{{$\varepsilon(H_0)$}}\\
\hline
{Voyager}&{73}&{1.37\%}&{20.55\%}&{98.63\%}&{2.33\%}\\
{Network}&{76}&{1.32\%}&{19.74\%}&{98.68\%}&{1.71\%}\\
\hline\hline \label{tab5}
\end{tabular}}}
\end{table*}

In this work, we do not consider the detector coordinates in the antenna response functions, which can impact the GW localization and the dark siren analysis. 
Hence, we wish to show the impact of the detector coordinates in the antenna response functions on estimating $H_0$ in the dark siren analysis. Since the detailed coordinates of Voyager and NEMO are currently uncertain, we make an assumption that the coordinates of Voyager and NEMO are the same as those of CE in the US and CE in Australia (detailed coordinates can be found in e.g., ref.~\cite{Jin:2022qnj}), respectively. 

{In figure~\ref{fig12}, we show the distributions of $\sigma_{d_{\rm L}}/d_{\rm L}$ and $\Delta\Omega$ for simulated BBHs at $z<0.15$ for Voyager and the Voyager-NEMO network with and without considering the detector coordinates in the antenna response functions to show the localization ability of the 2.5G detectors. We can see that compared to the single Voyager, the Voyager-NEMO network provides a slight improvement in the measurements of $d_{\rm L}$ and $\Delta\Omega$ because NEMO is not sensitive to the BBH mergers. When the detector coordinates are taken into account in the analysis, the sky localization $\Delta\Omega$ is significantly enhanced, though the improvements in $d_{\rm L}$ are limited.}

In figure~\ref{fig13}, we show the one-dimensional marginalized probability distributions for $H_0$ with and without considering the detector coordinates in the antenna response functions. {As seen, the scenario without considering detector coordinates provides constraints similar to those with detector coordinates. Although accounting for detector coordinates should yield more accurate localizations of the GW sources, it results in decreases in SNRs, leading to fewer detectable GW sources (see tables~\ref{tab3} and \ref{tab5}). These effects offset each other in the estimation of cosmological parameters, resulting in similar constraints for both scenarios, with and without considering the detector coordinates.} Concretely, when considering the detector coordinates in the antenna response functions, $H_0$ can be constrained to precisions of {2.33\%} and {1.71\%} using Voyager and the Voyager-NEMO network, respectively.
In conclusion, worse constraints can be obtained if the detector coordinates are considered in the antenna response functions. This work shows the optimistic scenario of the 2.5G GW detectors in the cosmological parameter estimations.


\section{Conclusion}\label{sec4}
{In this work, we explore the potential of the 2.5G ground-based GW detectors by considering three detection strategies: Voyager, NEMO, and the Voyager-NEMO network, in constraining cosmological parameters, based on the 10-year observation. Our main analysis is based on the BNS merger rate $R_{\rm m}(0)=320$ $\rm {Gpc^{-3}\ yr^{-1}}$. Furthermore, the impact of uncertainty in the BNS merger rate on estimating cosmological parameters is also discussed.}
We consider the optimistic scenario in which EM counterparts (SGRBs) can be detected and the conservative scenario in which EM counterparts are not available to perform cosmological analysis. For the simulation of the bright siren data, we calculate the expected detection distributions of GW-SGRB events. We consider the THESEUS-like mission as the representative of the SGRB detector.
We consider three typical dark energy models, i.e., the $\Lambda$CDM, $w$CDM, and $w_0w_a$CDM models, to make the cosmological analysis. In the conservative scenario, we choose the dark siren events with $\mathrm{SNR}\geq 100$ combined with the simulated galaxy catalogs to estimate cosmological parameters in the $\Lambda$CDM model.

{The mock bright siren data from Voyager, NEMO, and the Voyager-NEMO network can potentially measure $H_0$ to precisions of 2.08\%, 6.44\%, and 1.63\%, respectively, but it is rather challenging to measure other cosmological parameters by solely using the mock 2.5G bright siren data.}
Fortunately, the cosmological parameter degeneracy orientations of bright sirens and CMB are different, and thus the combination of them could break the cosmological parameter degeneracies, especially in the dynamical dark energy models. 
{Moreover, compared to the CMB data, the improvements on $\Omega_{\rm m}$, $H_0$, and $w$ constraints using CMB+Network are 84.3\%, 85.8\%, and 82.7\%, respectively. For the $w_0w_a$CDM model, the joint CMB+Network data give $\sigma(w_0)=0.44$ and $\sigma(w_a)=1.6$.}

{For the mock 2.5G dark siren scenario, $H_0$ is potentially measured to precisions of {2.32\% and 1.70\%} using Voyager and the Voyager-NEMO network, respectively. When combining dark sirens with bright sirens, $H_0$ is expected to reach precisions of {1.83\% and 1.35\%} for Voyager and the Voyager-NEMO network, respectively, which are close to or better than the distance ladder measurement. In addition, the ability of the standard sirens from the Voyager-NEMO network to constrain cosmological parameters is much worse than those of the 3G GW detectors.}

{We discuss the impact of uncertainty in the BNS merger rate on estimating cosmological parameters. We find that the BNS merger rates can severely affect the cosmological parameter estimations (measurement precisions of $H_0$ from 7.59\% to 1.21\%). Even in the scenario of the lower BNS merger rate, the ability of 2.5G bright sirens to break the cosmological parameter degeneracies is strong in the $w$CDM model, but main contributions of CMB+Network in the $\Lambda$CDM model come from CMB.} {In addition, we also discuss the impact of the detector coordinates in the antenna response functions on estimating $H_0$ in the dark siren scenario.}
{We emphasize that the constraint results in this work are optimistic since we consider the simplest GW waveform model and make some assumptions in the dark siren analysis. A more robust analysis will be undertaken in future work.}

We can conclude that {(i) it is more helpful to consider the 2.5G detector network than the single 2.5G GW observatory for the cosmological research,} (ii) the bright sirens from the 2.5G GW detectors could provide precise measurements on the Hubble constant {(a potential 1.63\% measurement)}, but are poor at measuring other cosmological parameters, (iii) the bright sirens from the 2.5G GW detectors could effectively break the cosmological parameter degeneracies generated by the CMB data and thus the combination of them could improve the measurement precisions of cosmological parameters, especially in the dynamical dark energy models, and (iiii) {the mock dark sirens from the 2.5G GW detectors give a potential {1.70\%} measurement on the Hubble constant,} while the combination of the bright and dark sirens can measure the Hubble constant to a precision of {1.35\%}.
It is worth expecting that the Hubble tension could be resolved in the era of the 2.5G ground-based GW detectors.

\acknowledgments
We thank the anonymous referee for the useful comments.  We also thank Yue-Yan Dong for the fruitful discussions and Paul D. Lasky for providing the sensitivity curve of NEMO. This work was supported by the National SKA Program of China (Grants Nos. 2022SKA0110200 and 2022SKA0110203), the National Natural Science Foundation of China (Grants Nos. 11975072, 11875102, and 11835009), and the National 111 Project (Grant No. B16009).

\bibliography{nemo_jcap}{}

\providecommand{\href}[2]{#2}\begingroup\raggedright\begin{thebibliography}{100}

\bibitem{WMAP:2003elm}
{\scshape WMAP} collaboration, \emph{{First year Wilkinson Microwave Anisotropy
  Probe (WMAP) observations: Determination of cosmological parameters}},
  \href{https://doi.org/10.1086/377226}{\emph{Astrophys. J. Suppl.} {\bfseries
  148} (2003) 175} [\href{https://arxiv.org/abs/astro-ph/0302209}{{\ttfamily
  astro-ph/0302209}}].

\bibitem{WMAP:2003ivt}
{\scshape WMAP} collaboration, \emph{{First year Wilkinson Microwave Anisotropy
  Probe (WMAP) observations: Preliminary maps and basic results}},
  \href{https://doi.org/10.1086/377253}{\emph{Astrophys. J. Suppl.} {\bfseries
  148} (2003) 1} [\href{https://arxiv.org/abs/astro-ph/0302207}{{\ttfamily
  astro-ph/0302207}}].

\bibitem{Planck:2018vyg}
{\scshape Planck} collaboration, \emph{{Planck 2018 results. VI. Cosmological
  parameters}},
  \href{https://doi.org/10.1051/0004-6361/201833910}{\emph{Astron. Astrophys.}
  {\bfseries 641} (2020) A6}
  [\href{https://arxiv.org/abs/1807.06209}{{\ttfamily 1807.06209}}].

\bibitem{Riess:2021jrx}
A.G.~Riess et~al., \emph{{A Comprehensive Measurement of the Local Value of the
  Hubble Constant with 1 km s$^{-1}$ Mpc$^{-1}$ Uncertainty from the Hubble
  Space Telescope and the SH0ES Team}},
  \href{https://doi.org/10.3847/2041-8213/ac5c5b}{\emph{Astrophys. J. Lett.}
  {\bfseries 934} (2022) L7}
  [\href{https://arxiv.org/abs/2112.04510}{{\ttfamily 2112.04510}}].

\bibitem{Verde:2019ivm}
L.~Verde, T.~Treu and A.G.~Riess, \emph{{Tensions between the Early and the
  Late Universe}},
  \href{https://doi.org/10.1038/s41550-019-0902-0}{\emph{Nature Astron.}
  {\bfseries 3} (2019) 891} [\href{https://arxiv.org/abs/1907.10625}{{\ttfamily
  1907.10625}}].

\bibitem{Riess:2019qba}
A.G.~Riess, \emph{{The Expansion of the Universe is Faster than Expected}},
  \href{https://doi.org/10.1038/s42254-019-0137-0}{\emph{Nature Rev. Phys.}
  {\bfseries 2} (2019) 10} [\href{https://arxiv.org/abs/2001.03624}{{\ttfamily
  2001.03624}}].

\bibitem{DiValentino:2021izs}
E.~Di~Valentino, O.~Mena, S.~Pan, L.~Visinelli, W.~Yang, A.~Melchiorri et~al.,
  \emph{{In the realm of the Hubble tension\textemdash{}a review of
  solutions}}, \href{https://doi.org/10.1088/1361-6382/ac086d}{\emph{Class.
  Quant. Grav.} {\bfseries 38} (2021) 153001}
  [\href{https://arxiv.org/abs/2103.01183}{{\ttfamily 2103.01183}}].

\bibitem{Kamionkowski:2022pkx}
M.~Kamionkowski and A.G.~Riess, \emph{{The Hubble Tension and Early Dark
  Energy}}, {\emph{Ann. Rev. Nucl. Part. Sci.} {\bfseries 73} (2023) 153}
  [\href{https://arxiv.org/abs/2211.04492}{{\ttfamily 2211.04492}}].

\bibitem{Perivolaropoulos:2021jda}
L.~Perivolaropoulos and F.~Skara, \emph{{Challenges for
  \ensuremath{\Lambda}CDM: An update}},
  \href{https://doi.org/10.1016/j.newar.2022.101659}{\emph{New Astron. Rev.}
  {\bfseries 95} (2022) 101659}
  [\href{https://arxiv.org/abs/2105.05208}{{\ttfamily 2105.05208}}].

\bibitem{Abdalla:2022yfr}
E.~Abdalla et~al., \emph{{Cosmology intertwined: A review of the particle
  physics, astrophysics, and cosmology associated with the cosmological
  tensions and anomalies}},
  \href{https://doi.org/10.1016/j.jheap.2022.04.002}{\emph{JHEAp} {\bfseries
  34} (2022) 49} [\href{https://arxiv.org/abs/2203.06142}{{\ttfamily
  2203.06142}}].

\bibitem{Guo:2018ans}
R.-Y.~Guo, J.-F.~Zhang and X.~Zhang, \emph{{Can the $H_0$ tension be resolved
  in extensions to $\Lambda$CDM cosmology?}},
  \href{https://doi.org/10.1088/1475-7516/2019/02/054}{\emph{JCAP} {\bfseries
  02} (2019) 054} [\href{https://arxiv.org/abs/1809.02340}{{\ttfamily
  1809.02340}}].

\bibitem{Poulin:2018cxd}
V.~Poulin, T.L.~Smith, T.~Karwal and M.~Kamionkowski, \emph{{Early Dark Energy
  Can Resolve The Hubble Tension}},
  \href{https://doi.org/10.1103/PhysRevLett.122.221301}{\emph{Phys. Rev. Lett.}
  {\bfseries 122} (2019) 221301}
  [\href{https://arxiv.org/abs/1811.04083}{{\ttfamily 1811.04083}}].

\bibitem{Cai:2021wgv}
R.-G.~Cai, Z.-K.~Guo, L.~Li, S.-J.~Wang and W.-W.~Yu, \emph{{Chameleon dark
  energy can resolve the Hubble tension}},
  \href{https://doi.org/10.1103/PhysRevD.103.L121302}{\emph{Phys. Rev. D}
  {\bfseries 103} (2021) 121302}
  [\href{https://arxiv.org/abs/2102.02020}{{\ttfamily 2102.02020}}].

\bibitem{Gao:2021xnk}
L.-Y.~Gao, Z.-W.~Zhao, S.-S.~Xue and X.~Zhang, \emph{{Relieving the H 0 tension
  with a new interacting dark energy model}},
  \href{https://doi.org/10.1088/1475-7516/2021/07/005}{\emph{JCAP} {\bfseries
  07} (2021) 005} [\href{https://arxiv.org/abs/2101.10714}{{\ttfamily
  2101.10714}}].

\bibitem{Yang:2018euj}
W.~Yang, S.~Pan, E.~Di~Valentino, R.C.~Nunes, S.~Vagnozzi and D.F.~Mota,
  \emph{{Tale of stable interacting dark energy, observational signatures, and
  the $H_0$ tension}},
  \href{https://doi.org/10.1088/1475-7516/2018/09/019}{\emph{JCAP} {\bfseries
  09} (2018) 019} [\href{https://arxiv.org/abs/1805.08252}{{\ttfamily
  1805.08252}}].

\bibitem{DiValentino:2019jae}
E.~Di~Valentino, A.~Melchiorri, O.~Mena and S.~Vagnozzi, \emph{{Nonminimal dark
  sector physics and cosmological tensions}},
  \href{https://doi.org/10.1103/PhysRevD.101.063502}{\emph{Phys. Rev. D}
  {\bfseries 101} (2020) 063502}
  [\href{https://arxiv.org/abs/1910.09853}{{\ttfamily 1910.09853}}].

\bibitem{DiValentino:2020zio}
E.~Di~Valentino et~al., \emph{{Snowmass2021 - Letter of interest cosmology
  intertwined II: The hubble constant tension}},
  \href{https://doi.org/10.1016/j.astropartphys.2021.102605}{\emph{Astropart.
  Phys.} {\bfseries 131} (2021) 102605}
  [\href{https://arxiv.org/abs/2008.11284}{{\ttfamily 2008.11284}}].

\bibitem{Liu:2019awo}
M.~Liu, Z.~Huang, X.~Luo, H.~Miao, N.K.~Singh and L.~Huang, \emph{{Can
  Non-standard Recombination Resolve the Hubble Tension?}},
  \href{https://doi.org/10.1007/s11433-019-1509-5}{\emph{Sci. China Phys. Mech.
  Astron.} {\bfseries 63} (2020) 290405}
  [\href{https://arxiv.org/abs/1912.00190}{{\ttfamily 1912.00190}}].

\bibitem{Zhang:2019cww}
X.~Zhang and Q.-G.~Huang, \emph{{Measuring H$_{0}$ from low-z datasets}},
  \href{https://doi.org/10.1007/s11433-019-1504-8}{\emph{Sci. China Phys. Mech.
  Astron.} {\bfseries 63} (2020) 290402}
  [\href{https://arxiv.org/abs/1911.09439}{{\ttfamily 1911.09439}}].

\bibitem{Ding:2019mmw}
Q.~Ding, T.~Nakama and Y.~Wang, \emph{{A gigaparsec-scale local void and the
  Hubble tension}}, \href{https://doi.org/10.1007/s11433-020-1531-0}{\emph{Sci.
  China Phys. Mech. Astron.} {\bfseries 63} (2020) 290403}
  [\href{https://arxiv.org/abs/1912.12600}{{\ttfamily 1912.12600}}].

\bibitem{Li:2020tds}
H.~Li and X.~Zhang, \emph{{A novel method of measuring cosmological distances
  using broad-line regions of quasars}},
  \href{https://doi.org/10.1016/j.scib.2020.04.038}{\emph{Sci. Bull.}
  {\bfseries 65} (2020) 1419}
  [\href{https://arxiv.org/abs/2005.10458}{{\ttfamily 2005.10458}}].

\bibitem{Vagnozzi:2021tjv}
S.~Vagnozzi, F.~Pacucci and A.~Loeb, \emph{{Implications for the Hubble tension
  from the ages of the oldest astrophysical objects}},
  \href{https://doi.org/10.1016/j.jheap.2022.07.004}{\emph{JHEAp} {\bfseries
  36} (2022) 27} [\href{https://arxiv.org/abs/2105.10421}{{\ttfamily
  2105.10421}}].

\bibitem{Guo:2019dui}
R.-Y.~Guo, J.-F.~Zhang and X.~Zhang, \emph{{Inflation model selection revisited
  after a 1.91\% measurement of the Hubble constant}},
  \href{https://doi.org/10.1007/s11433-019-1514-0}{\emph{Sci. China Phys. Mech.
  Astron.} {\bfseries 63} (2020) 290406}
  [\href{https://arxiv.org/abs/1910.13944}{{\ttfamily 1910.13944}}].

\bibitem{Feng:2019jqa}
L.~Feng, D.-Z.~He, H.-L.~Li, J.-F.~Zhang and X.~Zhang, \emph{{Constraints on
  active and sterile neutrinos in an interacting dark energy cosmology}},
  \href{https://doi.org/10.1007/s11433-019-1511-8}{\emph{Sci. China Phys. Mech.
  Astron.} {\bfseries 63} (2020) 290404}
  [\href{https://arxiv.org/abs/1910.03872}{{\ttfamily 1910.03872}}].

\bibitem{Gao:2022ahg}
L.-Y.~Gao, S.-S.~Xue and X.~Zhang, \emph{{Dark energy and matter interacting
  scenario to relieve H $_{0}$ and S $_{8}$ tensions*}},
  \href{https://doi.org/10.1088/1674-1137/ad2b52}{\emph{Chin. Phys. C}
  {\bfseries 48} (2024) 051001}
  [\href{https://arxiv.org/abs/2212.13146}{{\ttfamily 2212.13146}}].

\bibitem{Cao:2021zpf}
M.-D.~Cao, J.~Zheng, J.-Z.~Qi, X.~Zhang and Z.-H.~Zhu, \emph{{A New Way to
  Explore Cosmological Tensions Using Gravitational Waves and Strong
  Gravitational Lensing}},
  \href{https://doi.org/10.3847/1538-4357/ac7ce4}{\emph{Astrophys. J.}
  {\bfseries 934} (2022) 108}
  [\href{https://arxiv.org/abs/2112.14564}{{\ttfamily 2112.14564}}].

\bibitem{Schutz:1986gp}
B.F.~Schutz, \emph{{Determining the Hubble Constant from Gravitational Wave
  Observations}}, \href{https://doi.org/10.1038/323310a0}{\emph{Nature}
  {\bfseries 323} (1986) 310}.

\bibitem{Holz:2005df}
D.E.~Holz and S.A.~Hughes, \emph{{Using gravitational-wave standard sirens}},
  \href{https://doi.org/10.1086/431341}{\emph{Astrophys. J.} {\bfseries 629}
  (2005) 15} [\href{https://arxiv.org/abs/astro-ph/0504616}{{\ttfamily
  astro-ph/0504616}}].

\bibitem{Nissanke:2009kt}
S.~Nissanke, D.E.~Holz, S.A.~Hughes, N.~Dalal and J.L.~Sievers,
  \emph{{Exploring short gamma-ray bursts as gravitational-wave standard
  sirens}}, \href{https://doi.org/10.1088/0004-637X/725/1/496}{\emph{Astrophys.
  J.} {\bfseries 725} (2010) 496}
  [\href{https://arxiv.org/abs/0904.1017}{{\ttfamily 0904.1017}}].

\bibitem{Dalal:2006qt}
N.~Dalal, D.E.~Holz, S.A.~Hughes and B.~Jain, \emph{{Short grb and binary black
  hole standard sirens as a probe of dark energy}},
  \href{https://doi.org/10.1103/PhysRevD.74.063006}{\emph{Phys. Rev. D}
  {\bfseries 74} (2006) 063006}
  [\href{https://arxiv.org/abs/astro-ph/0601275}{{\ttfamily
  astro-ph/0601275}}].

\bibitem{Chen:2017rfc}
H.-Y.~Chen, M.~Fishbach and D.E.~Holz, \emph{{A two per cent Hubble constant
  measurement from standard sirens within five years}},
  \href{https://doi.org/10.1038/s41586-018-0606-0}{\emph{Nature} {\bfseries
  562} (2018) 545} [\href{https://arxiv.org/abs/1712.06531}{{\ttfamily
  1712.06531}}].

\bibitem{Cutler:2009qv}
C.~Cutler and D.E.~Holz, \emph{{Ultra-high precision cosmology from
  gravitational waves}},
  \href{https://doi.org/10.1103/PhysRevD.80.104009}{\emph{Phys. Rev. D}
  {\bfseries 80} (2009) 104009}
  [\href{https://arxiv.org/abs/0906.3752}{{\ttfamily 0906.3752}}].

\bibitem{Cai:2016sby}
R.-G.~Cai and T.~Yang, \emph{{Estimating cosmological parameters by the
  simulated data of gravitational waves from the Einstein Telescope}},
  \href{https://doi.org/10.1103/PhysRevD.95.044024}{\emph{Phys. Rev. D}
  {\bfseries 95} (2017) 044024}
  [\href{https://arxiv.org/abs/1608.08008}{{\ttfamily 1608.08008}}].

\bibitem{Cai:2017cbj}
R.-G.~Cai, Z.~Cao, Z.-K.~Guo, S.-J.~Wang and T.~Yang, \emph{{The
  Gravitational-Wave Physics}},
  \href{https://doi.org/10.1093/nsr/nwx029}{\emph{Natl. Sci. Rev.} {\bfseries
  4} (2017) 687} [\href{https://arxiv.org/abs/1703.00187}{{\ttfamily
  1703.00187}}].

\bibitem{Gray:2019ksv}
R.~Gray et~al., \emph{{Cosmological inference using gravitational wave standard
  sirens: A mock data analysis}},
  \href{https://doi.org/10.1103/PhysRevD.101.122001}{\emph{Phys. Rev. D}
  {\bfseries 101} (2020) 122001}
  [\href{https://arxiv.org/abs/1908.06050}{{\ttfamily 1908.06050}}].

\bibitem{Vitale:2018wlg}
S.~Vitale and H.-Y.~Chen, \emph{{Measuring the Hubble constant with neutron
  star black hole mergers}},
  \href{https://doi.org/10.1103/PhysRevLett.121.021303}{\emph{Phys. Rev. Lett.}
  {\bfseries 121} (2018) 021303}
  [\href{https://arxiv.org/abs/1804.07337}{{\ttfamily 1804.07337}}].

\bibitem{Lagos:2019kds}
M.~Lagos, M.~Fishbach, P.~Landry and D.E.~Holz, \emph{{Standard sirens with a
  running Planck mass}},
  \href{https://doi.org/10.1103/PhysRevD.99.083504}{\emph{Phys. Rev. D}
  {\bfseries 99} (2019) 083504}
  [\href{https://arxiv.org/abs/1901.03321}{{\ttfamily 1901.03321}}].

\bibitem{Camera:2013xfa}
S.~Camera and A.~Nishizawa, \emph{{Beyond Concordance Cosmology with
  Magnification of Gravitational-Wave Standard Sirens}},
  \href{https://doi.org/10.1103/PhysRevLett.110.151103}{\emph{Phys. Rev. Lett.}
  {\bfseries 110} (2013) 151103}
  [\href{https://arxiv.org/abs/1303.5446}{{\ttfamily 1303.5446}}].

\bibitem{Mukherjee:2019qmm}
S.~Mukherjee, G.~Lavaux, F.R.~Bouchet, J.~Jasche, B.D.~Wandelt, S.M.~Nissanke
  et~al., \emph{{Velocity correction for Hubble constant measurements from
  standard sirens}},
  \href{https://doi.org/10.1051/0004-6361/201936724}{\emph{Astron. Astrophys.}
  {\bfseries 646} (2021) A65}
  [\href{https://arxiv.org/abs/1909.08627}{{\ttfamily 1909.08627}}].

\bibitem{DAgostino:2019hvh}
R.~D'Agostino and R.C.~Nunes, \emph{{Probing observational bounds on
  scalar-tensor theories from standard sirens}},
  \href{https://doi.org/10.1103/PhysRevD.100.044041}{\emph{Phys. Rev. D}
  {\bfseries 100} (2019) 044041}
  [\href{https://arxiv.org/abs/1907.05516}{{\ttfamily 1907.05516}}].

\bibitem{Wang:2018lun}
L.-F.~Wang, X.-N.~Zhang, J.-F.~Zhang and X.~Zhang, \emph{{Impacts of
  gravitational-wave standard siren observation of the Einstein Telescope on
  weighing neutrinos in cosmology}},
  \href{https://doi.org/10.1016/j.physletb.2018.05.027}{\emph{Phys. Lett. B}
  {\bfseries 782} (2018) 87}
  [\href{https://arxiv.org/abs/1802.04720}{{\ttfamily 1802.04720}}].

\bibitem{Cai:2017aea}
R.-G.~Cai, T.-B.~Liu, X.-W.~Liu, S.-J.~Wang and T.~Yang, \emph{{Probing cosmic
  anisotropy with gravitational waves as standard sirens}},
  \href{https://doi.org/10.1103/PhysRevD.97.103005}{\emph{Phys. Rev. D}
  {\bfseries 97} (2018) 103005}
  [\href{https://arxiv.org/abs/1712.00952}{{\ttfamily 1712.00952}}].

\bibitem{Jin:2020hmc}
S.-J.~Jin, D.-Z.~He, Y.~Xu, J.-F.~Zhang and X.~Zhang, \emph{{Forecast for
  cosmological parameter estimation with gravitational-wave standard siren
  observation from the Cosmic Explorer}},
  \href{https://doi.org/10.1088/1475-7516/2020/03/051}{\emph{JCAP} {\bfseries
  03} (2020) 051} [\href{https://arxiv.org/abs/2001.05393}{{\ttfamily
  2001.05393}}].

\bibitem{Belgacem:2019tbw}
E.~Belgacem, Y.~Dirian, S.~Foffa, E.J.~Howell, M.~Maggiore and T.~Regimbau,
  \emph{{Cosmology and dark energy from joint gravitational wave-GRB
  observations}},
  \href{https://doi.org/10.1088/1475-7516/2019/08/015}{\emph{JCAP} {\bfseries
  08} (2019) 015} [\href{https://arxiv.org/abs/1907.01487}{{\ttfamily
  1907.01487}}].

\bibitem{Howlett:2019mdh}
C.~Howlett and T.M.~Davis, \emph{{Standard siren speeds: improving velocities
  in gravitational-wave measurements of $H_0$}},
  \href{https://doi.org/10.1093/mnras/staa049}{\emph{Mon. Not. Roy. Astron.
  Soc.} {\bfseries 492} (2020) 3803}
  [\href{https://arxiv.org/abs/1909.00587}{{\ttfamily 1909.00587}}].

\bibitem{Zhang:2019loq}
J.-F.~Zhang, M.~Zhang, S.-J.~Jin, J.-Z.~Qi and X.~Zhang, \emph{{Cosmological
  parameter estimation with future gravitational wave standard siren
  observation from the Einstein Telescope}},
  \href{https://doi.org/10.1088/1475-7516/2019/09/068}{\emph{JCAP} {\bfseries
  09} (2019) 068} [\href{https://arxiv.org/abs/1907.03238}{{\ttfamily
  1907.03238}}].

\bibitem{Wu:2022dgy}
P.-J.~Wu, Y.~Shao, S.-J.~Jin and X.~Zhang, \emph{{A path to precision
  cosmology: synergy between four promising late-universe cosmological
  probes}}, \href{https://doi.org/10.1088/1475-7516/2023/06/052}{\emph{JCAP}
  {\bfseries 06} (2023) 052}
  [\href{https://arxiv.org/abs/2202.09726}{{\ttfamily 2202.09726}}].

\bibitem{Ezquiaga:2022zkx}
J.M.~Ezquiaga and D.E.~Holz, \emph{{Spectral Sirens: Cosmology from the Full
  Mass Distribution of Compact Binaries}},
  \href{https://doi.org/10.1103/PhysRevLett.129.061102}{\emph{Phys. Rev. Lett.}
  {\bfseries 129} (2022) 061102}
  [\href{https://arxiv.org/abs/2202.08240}{{\ttfamily 2202.08240}}].

\bibitem{Wang:2019tto}
L.-F.~Wang, Z.-W.~Zhao, J.-F.~Zhang and X.~Zhang, \emph{{A preliminary forecast
  for cosmological parameter estimation with gravitational-wave standard sirens
  from TianQin}},
  \href{https://doi.org/10.1088/1475-7516/2020/11/012}{\emph{JCAP} {\bfseries
  11} (2020) 012} [\href{https://arxiv.org/abs/1907.01838}{{\ttfamily
  1907.01838}}].

\bibitem{Zhao:2019gyk}
Z.-W.~Zhao, L.-F.~Wang, J.-F.~Zhang and X.~Zhang, \emph{{Prospects for
  improving cosmological parameter estimation with gravitational-wave standard
  sirens from Taiji}},
  \href{https://doi.org/10.1016/j.scib.2020.04.032}{\emph{Sci. Bull.}
  {\bfseries 65} (2020) 1340}
  [\href{https://arxiv.org/abs/1912.11629}{{\ttfamily 1912.11629}}].

\bibitem{Wang:2021srv}
L.-F.~Wang, S.-J.~Jin, J.-F.~Zhang and X.~Zhang, \emph{{Forecast for
  cosmological parameter estimation with gravitational-wave standard sirens
  from the LISA-Taiji network}},
  \href{https://doi.org/10.1007/s11433-021-1736-6}{\emph{Sci. China Phys. Mech.
  Astron.} {\bfseries 65} (2022) 210411}
  [\href{https://arxiv.org/abs/2101.11882}{{\ttfamily 2101.11882}}].

\bibitem{Jin:2021pcv}
S.-J.~Jin, L.-F.~Wang, P.-J.~Wu, J.-F.~Zhang and X.~Zhang, \emph{{How can
  gravitational-wave standard sirens and 21-cm intensity mapping jointly
  provide a precise late-universe cosmological probe?}},
  \href{https://doi.org/10.1103/PhysRevD.104.103507}{\emph{Phys. Rev. D}
  {\bfseries 104} (2021) 103507}
  [\href{https://arxiv.org/abs/2106.01859}{{\ttfamily 2106.01859}}].

\bibitem{Yu:2021nvx}
J.~Yu, H.~Song, S.~Ai, H.~Gao, F.~Wang, Y.~Wang et~al., \emph{{Multimessenger
  Detection Rates and Distributions of Binary Neutron Star Mergers and Their
  Cosmological Implications}},
  \href{https://doi.org/10.3847/1538-4357/ac0628}{\emph{Astrophys. J.}
  {\bfseries 916} (2021) 54}
  [\href{https://arxiv.org/abs/2104.12374}{{\ttfamily 2104.12374}}].

\bibitem{Chen:2023dgw}
H.-Y.~Chen, C.~Talbot and E.A.~Chase, \emph{{Mitigating the Counterpart
  Selection Effect for Standard Sirens}},
  \href{https://doi.org/10.1103/PhysRevLett.132.191003}{\emph{Phys. Rev. Lett.}
  {\bfseries 132} (2024) 191003}
  [\href{https://arxiv.org/abs/2307.10402}{{\ttfamily 2307.10402}}].

\bibitem{LISACosmologyWorkingGroup:2022jok}
{\scshape LISA Cosmology Working Group} collaboration, \emph{{Cosmology with
  the Laser Interferometer Space Antenna}},
  \href{https://doi.org/10.1007/s41114-023-00045-2}{\emph{Living Rev. Rel.}
  {\bfseries 26} (2023) 5} [\href{https://arxiv.org/abs/2204.05434}{{\ttfamily
  2204.05434}}].

\bibitem{Dhani:2022ulg}
A.~Dhani, S.~Borhanian, A.~Gupta and B.~Sathyaprakash, \emph{{Cosmography with
  bright and Love sirens}},  \href{https://arxiv.org/abs/2212.13183}{{\ttfamily
  2212.13183}}.

\bibitem{Zhu:2021bpp}
L.-G.~Zhu, L.-H.~Xie, Y.-M.~Hu, S.~Liu, E.-K.~Li, N.R.~Napolitano et~al.,
  \emph{{Constraining the Hubble constant to a precision of about 1\% using
  multi-band dark standard siren detections}},
  \href{https://doi.org/10.1007/s11433-021-1859-9}{\emph{Sci. China Phys. Mech.
  Astron.} {\bfseries 65} (2022) 259811}
  [\href{https://arxiv.org/abs/2110.05224}{{\ttfamily 2110.05224}}].

\bibitem{Qi:2021iic}
J.-Z.~Qi, S.-J.~Jin, X.-L.~Fan, J.-F.~Zhang and X.~Zhang, \emph{{Using a
  multi-messenger and multi-wavelength observational strategy to probe the
  nature of dark energy through direct measurements of cosmic expansion
  history}}, \href{https://doi.org/10.1088/1475-7516/2021/12/042}{\emph{JCAP}
  {\bfseries 12} (2021) 042}
  [\href{https://arxiv.org/abs/2102.01292}{{\ttfamily 2102.01292}}].

\bibitem{Jin:2022tdf}
S.-J.~Jin, R.-Q.~Zhu, L.-F.~Wang, H.-L.~Li, J.-F.~Zhang and X.~Zhang,
  \emph{{Impacts of gravitational-wave standard siren observations from
  Einstein Telescope and Cosmic Explorer on weighing neutrinos in interacting
  dark energy models}},
  \href{https://doi.org/10.1088/1572-9494/ac7b76}{\emph{Commun. Theor. Phys.}
  {\bfseries 74} (2022) 105404}
  [\href{https://arxiv.org/abs/2204.04689}{{\ttfamily 2204.04689}}].

\bibitem{Wang:2022oou}
L.-F.~Wang, Y.~Shao, J.-F.~Zhang and X.~Zhang, \emph{{Ultra-low-frequency
  gravitational waves from individual supermassive black hole binaries as
  standard sirens}},  \href{https://arxiv.org/abs/2201.00607}{{\ttfamily
  2201.00607}}.

\bibitem{Song:2022siz}
J.-Y.~Song, L.-F.~Wang, Y.~Li, Z.-W.~Zhao, J.-F.~Zhang, W.~Zhao et~al.,
  \emph{{Synergy between CSST galaxy survey and gravitational-wave observation:
  Inferring the Hubble constant from dark standard sirens}},
  \href{https://doi.org/10.1007/s11433-023-2260-2}{\emph{Sci. China Phys. Mech.
  Astron.} {\bfseries 67} (2024) 230411}
  [\href{https://arxiv.org/abs/2212.00531}{{\ttfamily 2212.00531}}].

\bibitem{Jin:2023zhi}
S.-J.~Jin, S.-S.~Xing, Y.~Shao, J.-F.~Zhang and X.~Zhang, \emph{{Joint
  constraints on cosmological parameters using future multi-band gravitational
  wave standard siren observations*}},
  \href{https://doi.org/10.1088/1674-1137/acc8be}{\emph{Chin. Phys. C}
  {\bfseries 47} (2023) 065104}
  [\href{https://arxiv.org/abs/2301.06722}{{\ttfamily 2301.06722}}].

\bibitem{Hou:2022rvk}
W.-T.~Hou, J.-Z.~Qi, T.~Han, J.-F.~Zhang, S.~Cao and X.~Zhang, \emph{{Prospects
  for constraining interacting dark energy models from gravitational wave and
  gamma ray burst joint observation}},
  \href{https://doi.org/10.1088/1475-7516/2023/05/017}{\emph{JCAP} {\bfseries
  05} (2023) 017} [\href{https://arxiv.org/abs/2211.10087}{{\ttfamily
  2211.10087}}].

\bibitem{Borhanian:2020vyr}
S.~Borhanian, A.~Dhani, A.~Gupta, K.G.~Arun and B.S.~Sathyaprakash, \emph{{Dark
  Sirens to Resolve the Hubble\textendash{}Lema\^\i{}tre Tension}},
  \href{https://doi.org/10.3847/2041-8213/abcaf5}{\emph{Astrophys. J. Lett.}
  {\bfseries 905} (2020) L28}
  [\href{https://arxiv.org/abs/2007.02883}{{\ttfamily 2007.02883}}].

\bibitem{Jin:2023sfc}
S.-J.~Jin, Y.-Z.~Zhang, J.-Y.~Song, J.-F.~Zhang and X.~Zhang,
  \emph{{Taiji-TianQin-LISA network: Precisely measuring the Hubble constant
  using both bright and dark sirens}},
  \href{https://doi.org/10.1007/s11433-023-2276-1}{\emph{Sci. China Phys. Mech.
  Astron.} {\bfseries 67} (2024) 220412}
  [\href{https://arxiv.org/abs/2305.19714}{{\ttfamily 2305.19714}}].

\bibitem{Jin:2022qnj}
S.-J.~Jin, T.-N.~Li, J.-F.~Zhang and X.~Zhang, \emph{{Prospects for measuring
  the Hubble constant and dark energy using gravitational-wave dark sirens with
  neutron star tidal deformation}},
  \href{https://doi.org/10.1088/1475-7516/2023/08/070}{\emph{JCAP} {\bfseries
  08} (2023) 070} [\href{https://arxiv.org/abs/2202.11882}{{\ttfamily
  2202.11882}}].

\bibitem{Zhu:2023jti}
L.-G.~Zhu and X.~Chen, \emph{{The Dark Side of Using Dark Sirens to Constrain
  the Hubble\textendash{}Lema\^\i{}tre Constant}},
  \href{https://doi.org/10.3847/1538-4357/acc24b}{\emph{Astrophys. J.}
  {\bfseries 948} (2023) 26}
  [\href{https://arxiv.org/abs/2302.10621}{{\ttfamily 2302.10621}}].

\bibitem{Han:2023exn}
T.~Han, S.-J.~Jin, J.-F.~Zhang and X.~Zhang, \emph{{A comprehensive forecast
  for cosmological parameter estimation using joint observations of
  gravitational waves and short $\gamma $-ray bursts}},
  \href{https://doi.org/10.1140/epjc/s10052-024-12999-w}{\emph{Eur. Phys. J. C}
  {\bfseries 84} (2024) 663}
  [\href{https://arxiv.org/abs/2309.14965}{{\ttfamily 2309.14965}}].

\bibitem{Li:2023gtu}
T.-N.~Li, S.-J.~Jin, H.-L.~Li, J.-F.~Zhang and X.~Zhang, \emph{{Prospects for
  Probing the Interaction between Dark Energy and Dark Matter Using
  Gravitational-wave Dark Sirens with Neutron Star Tidal Deformation}},
  \href{https://doi.org/10.3847/1538-4357/ad1bc9}{\emph{Astrophys. J.}
  {\bfseries 963} (2024) 52}
  [\href{https://arxiv.org/abs/2310.15879}{{\ttfamily 2310.15879}}].

\bibitem{Muttoni:2023prw}
N.~Muttoni, D.~Laghi, N.~Tamanini, S.~Marsat and D.~Izquierdo-Villalba,
  \emph{{Dark siren cosmology with binary black holes in the era of
  third-generation gravitational wave detectors}},
  \href{https://doi.org/10.1103/PhysRevD.108.043543}{\emph{Phys. Rev. D}
  {\bfseries 108} (2023) 043543}
  [\href{https://arxiv.org/abs/2303.10693}{{\ttfamily 2303.10693}}].

\bibitem{Branchesi:2023mws}
M.~Branchesi et~al., \emph{{Science with the Einstein Telescope: a comparison
  of different designs}},
  \href{https://doi.org/10.1088/1475-7516/2023/07/068}{\emph{JCAP} {\bfseries
  07} (2023) 068} [\href{https://arxiv.org/abs/2303.15923}{{\ttfamily
  2303.15923}}].

\bibitem{Zheng:2024mbo}
J.~Zheng, X.-H.~Liu and J.-Z.~Qi, \emph{{Joint observations of late universe
  probes: cosmological parameter constraints from gravitational wave and Type
  Ia supernova data}},  \href{https://arxiv.org/abs/2407.05686}{{\ttfamily
  2407.05686}}.

\bibitem{Dong:2024bvw}
Y.-Y.~Dong, J.-Y.~Song, S.-J.~Jin, J.-F.~Zhang and X.~Zhang, \emph{{Enhancing
  dark siren cosmology through multi-band gravitational wave synergetic
  observations}},  \href{https://arxiv.org/abs/2404.18188}{{\ttfamily
  2404.18188}}.

\bibitem{Zheng:2022gfi}
J.~Zheng, Y.~Chen, T.~Xu and Z.-H.~Zhu, \emph{{Investigating the dynamical
  models of cosmology with recent observations and upcoming gravitational-wave
  data}}, \href{https://doi.org/10.1140/epjp/s13360-022-02718-3}{\emph{Eur.
  Phys. J. Plus} {\bfseries 137} (2022) 509}
  [\href{https://arxiv.org/abs/2201.07011}{{\ttfamily 2201.07011}}].

\bibitem{Feng:2024lzh}
L.~Feng, T.~Han, J.-F.~Zhang and X.~Zhang, \emph{{Prospects for weighing
  neutrinos in interacting dark energy models using joint observations of
  gravitational waves and $\gamma$-ray bursts}},
  \href{https://doi.org/10.1088/1674-1137/ad5ae4}{\emph{Chin. Phys. C}
  {\bfseries 48} (2024) 095104}
  [\href{https://arxiv.org/abs/2404.19530}{{\ttfamily 2404.19530}}].

\bibitem{Yu:2023ico}
J.~Yu, Z.~Liu, X.~Yang, Y.~Wang, P.~Zhang, X.~Zhang et~al., \emph{{Measuring
  the Hubble Constant of Binary Neutron Star and Neutron Star\textendash{}Black
  Hole Coalescences: Bright Sirens and Dark Sirens}},
  \href{https://doi.org/10.3847/1538-4365/ad0ece}{\emph{Astrophys. J. Suppl.}
  {\bfseries 270} (2024) 24}
  [\href{https://arxiv.org/abs/2311.11588}{{\ttfamily 2311.11588}}].

\bibitem{Xiao:2024nmi}
S.-R.~Xiao, Y.~Shao, L.-F.~Wang, J.-Y.~Song, L.~Feng, J.-F.~Zhang et~al.,
  \emph{{Nanohertz gravitational waves from a quasar-based supermassive black
  hole binary population model as dark sirens}},
  \href{https://arxiv.org/abs/2408.00609}{{\ttfamily 2408.00609}}.

\bibitem{LIGOScientific:2017vwq}
{\scshape LIGO Scientific, Virgo} collaboration, \emph{{GW170817: Observation
  of Gravitational Waves from a Binary Neutron Star Inspiral}},
  \href{https://doi.org/10.1103/PhysRevLett.119.161101}{\emph{Phys. Rev. Lett.}
  {\bfseries 119} (2017) 161101}
  [\href{https://arxiv.org/abs/1710.05832}{{\ttfamily 1710.05832}}].

\bibitem{LIGOScientific:2017ync}
{\scshape LIGO Scientific, Virgo, Fermi GBM, INTEGRAL, IceCube, AstroSat
  Cadmium Zinc Telluride Imager Team, IPN, Insight-Hxmt, ANTARES, Swift, AGILE
  Team, 1M2H Team, Dark Energy Camera GW-EM, DES, DLT40, GRAWITA, Fermi-LAT,
  ATCA, ASKAP, Las Cumbres Observatory Group, OzGrav, DWF (Deeper Wider Faster
  Program), AST3, CAASTRO, VINROUGE, MASTER, J-GEM, GROWTH, JAGWAR,
  CaltechNRAO, TTU-NRAO, NuSTAR, Pan-STARRS, MAXI Team, TZAC Consortium, KU,
  Nordic Optical Telescope, ePESSTO, GROND, Texas Tech University, SALT Group,
  TOROS, BOOTES, MWA, CALET, IKI-GW Follow-up, H.E.S.S., LOFAR, LWA, HAWC,
  Pierre Auger, ALMA, Euro VLBI Team, Pi of Sky, Chandra Team at McGill
  University, DFN, ATLAS Telescopes, High Time Resolution Universe Survey,
  RIMAS, RATIR, SKA South Africa/MeerKAT} collaboration, \emph{{Multi-messenger
  Observations of a Binary Neutron Star Merger}},
  \href{https://doi.org/10.3847/2041-8213/aa91c9}{\emph{Astrophys. J. Lett.}
  {\bfseries 848} (2017) L12}
  [\href{https://arxiv.org/abs/1710.05833}{{\ttfamily 1710.05833}}].

\bibitem{LIGOScientific:2017adf}
{\scshape LIGO Scientific, Virgo, 1M2H, Dark Energy Camera GW-E, DES, DLT40,
  Las Cumbres Observatory, VINROUGE, MASTER} collaboration, \emph{{A
  gravitational-wave standard siren measurement of the Hubble constant}},
  \href{https://doi.org/10.1038/nature24471}{\emph{Nature} {\bfseries 551}
  (2017) 85} [\href{https://arxiv.org/abs/1710.05835}{{\ttfamily 1710.05835}}].

\bibitem{LIGOScientific:2021aug}
{\scshape LIGO Scientific, Virgo,, KAGRA, VIRGO} collaboration,
  \emph{{Constraints on the Cosmic Expansion History from GWTC\textendash{}3}},
  \href{https://doi.org/10.3847/1538-4357/ac74bb}{\emph{Astrophys. J.}
  {\bfseries 949} (2023) 76}
  [\href{https://arxiv.org/abs/2111.03604}{{\ttfamily 2111.03604}}].

\bibitem{LIGO:2020xsf}
{\scshape LIGO} collaboration, \emph{{A cryogenic silicon interferometer for
  gravitational-wave detection}},
  \href{https://doi.org/10.1088/1361-6382/ab9143}{\emph{Class. Quant. Grav.}
  {\bfseries 37} (2020) 165003}
  [\href{https://arxiv.org/abs/2001.11173}{{\ttfamily 2001.11173}}].

\bibitem{Ackley:2020atn}
K.~Ackley et~al., \emph{{Neutron Star Extreme Matter Observatory: A
  kilohertz-band gravitational-wave detector in the global network}},
  \href{https://doi.org/10.1017/pasa.2020.39}{\emph{Publ. Astron. Soc.
  Austral.} {\bfseries 37} (2020) e047}
  [\href{https://arxiv.org/abs/2007.03128}{{\ttfamily 2007.03128}}].

\bibitem{Evans:2021gyd}
M.~Evans et~al., \emph{{A Horizon Study for Cosmic Explorer: Science,
  Observatories, and Community}},
  \href{https://arxiv.org/abs/2109.09882}{{\ttfamily 2109.09882}}.

\bibitem{Sarin:2021qqo}
N.~Sarin and P.D.~Lasky, \emph{{Multimessenger astronomy with a kHz-band
  gravitational-wave observatory}},
  \href{https://doi.org/10.1017/pasa.2022.1}{\emph{Publ. Astron. Soc. Austral.}
  {\bfseries 39} (2022) e007}
  [\href{https://arxiv.org/abs/2110.10892}{{\ttfamily 2110.10892}}].

\bibitem{Ballmer:2022uxx}
S.W.~Ballmer et~al., \emph{{Snowmass2021 Cosmic Frontier White Paper: Future
  Gravitational-Wave Detector Facilities}},  in \emph{{Snowmass 2021}}, 3, 2022
  [\href{https://arxiv.org/abs/2203.08228}{{\ttfamily 2203.08228}}].

\bibitem{Relton:2021cax}
P.~Relton and V.~Raymond, \emph{{Parameter estimation bias from overlapping
  binary black hole events in second generation interferometers}},
  \href{https://doi.org/10.1103/PhysRevD.104.084039}{\emph{Phys. Rev. D}
  {\bfseries 104} (2021) 084039}
  [\href{https://arxiv.org/abs/2103.16225}{{\ttfamily 2103.16225}}].

\bibitem{Borhanian:2022czq}
S.~Borhanian and B.S.~Sathyaprakash, \emph{{Listening to the Universe with Next
  Generation Ground-Based Gravitational-Wave Detectors}},
  \href{https://arxiv.org/abs/2202.11048}{{\ttfamily 2202.11048}}.

\bibitem{Madau:2014bja}
P.~Madau and M.~Dickinson, \emph{{Cosmic Star Formation History}},
  \href{https://doi.org/10.1146/annurev-astro-081811-125615}{\emph{Ann. Rev.
  Astron. Astrophys.} {\bfseries 52} (2014) 415}
  [\href{https://arxiv.org/abs/1403.0007}{{\ttfamily 1403.0007}}].

\bibitem{Vitale:2018yhm}
S.~Vitale, W.M.~Farr, K.~Ng and C.L.~Rodriguez, \emph{{Measuring the star
  formation rate with gravitational waves from binary black holes}},
  \href{https://doi.org/10.3847/2041-8213/ab50c0}{\emph{Astrophys. J. Lett.}
  {\bfseries 886} (2019) L1}
  [\href{https://arxiv.org/abs/1808.00901}{{\ttfamily 1808.00901}}].

\bibitem{KAGRA:2021duu}
{\scshape KAGRA, VIRGO, LIGO Scientific} collaboration, \emph{{Population of
  Merging Compact Binaries Inferred Using Gravitational Waves through GWTC-3}},
  \href{https://doi.org/10.1103/PhysRevX.13.011048}{\emph{Phys. Rev. X}
  {\bfseries 13} (2023) 011048}
  [\href{https://arxiv.org/abs/2111.03634}{{\ttfamily 2111.03634}}].

\bibitem{LIGOScientific:2020ibl}
{\scshape LIGO Scientific, Virgo} collaboration, \emph{{GWTC-2: Compact Binary
  Coalescences Observed by LIGO and Virgo During the First Half of the Third
  Observing Run}},
  \href{https://doi.org/10.1103/PhysRevX.11.021053}{\emph{Phys. Rev. X}
  {\bfseries 11} (2021) 021053}
  [\href{https://arxiv.org/abs/2010.14527}{{\ttfamily 2010.14527}}].

\bibitem{LIGOScientific:2020kqk}
{\scshape LIGO Scientific, Virgo} collaboration, \emph{{Population Properties
  of Compact Objects from the Second LIGO-Virgo Gravitational-Wave Transient
  Catalog}}, \href{https://doi.org/10.3847/2041-8213/abe949}{\emph{Astrophys.
  J. Lett.} {\bfseries 913} (2021) L7}
  [\href{https://arxiv.org/abs/2010.14533}{{\ttfamily 2010.14533}}].

\bibitem{Cho:2015dra}
H.-S.~Cho, \emph{{Parameter estimation using a complete signal and inspiral
  templates for low mass binary black holes with Advanced LIGO sensitivity}},
  \href{https://doi.org/10.1088/0264-9381/32/23/235007}{\emph{Class. Quant.
  Grav.} {\bfseries 32} (2015) 235007}
  [\href{https://arxiv.org/abs/1502.04399}{{\ttfamily 1502.04399}}].

\bibitem{Kumar:2016dhh}
P.~Kumar, T.~Chu, H.~Fong, H.P.~Pfeiffer, M.~Boyle, D.A.~Hemberger et~al.,
  \emph{{Accuracy of binary black hole waveform models for aligned-spin
  binaries}}, \href{https://doi.org/10.1103/PhysRevD.93.104050}{\emph{Phys.
  Rev. D} {\bfseries 93} (2016) 104050}
  [\href{https://arxiv.org/abs/1601.05396}{{\ttfamily 1601.05396}}].

\bibitem{Sathyaprakash:2009xs}
B.S.~Sathyaprakash and B.F.~Schutz, \emph{{Physics, Astrophysics and Cosmology
  with Gravitational Waves}},
  \href{https://doi.org/10.12942/lrr-2009-2}{\emph{Living Rev. Rel.} {\bfseries
  12} (2009) 2} [\href{https://arxiv.org/abs/0903.0338}{{\ttfamily
  0903.0338}}].

\bibitem{Blanchet:2004bb}
L.~Blanchet and B.R.~Iyer, \emph{{Hadamard regularization of the third
  post-Newtonian gravitational wave generation of two point masses}},
  \href{https://doi.org/10.1103/PhysRevD.71.024004}{\emph{Phys. Rev. D}
  {\bfseries 71} (2005) 024004}
  [\href{https://arxiv.org/abs/gr-qc/0409094}{{\ttfamily gr-qc/0409094}}].

\bibitem{Arun:2004hn}
K.G.~Arun, B.R.~Iyer, B.S.~Sathyaprakash and P.A.~Sundararajan,
  \emph{{Parameter estimation of inspiralling compact binaries using 3.5
  post-Newtonian gravitational wave phasing: The Non-spinning case}},
  \href{https://doi.org/10.1103/PhysRevD.71.084008}{\emph{Phys. Rev. D}
  {\bfseries 71} (2005) 084008}
  [\href{https://arxiv.org/abs/gr-qc/0411146}{{\ttfamily gr-qc/0411146}}].

\bibitem{Robson:2018ifk}
T.~Robson, N.J.~Cornish and C.~Liu, \emph{{The construction and use of LISA
  sensitivity curves}},
  \href{https://doi.org/10.1088/1361-6382/ab1101}{\emph{Class. Quant. Grav.}
  {\bfseries 36} (2019) 105011}
  [\href{https://arxiv.org/abs/1803.01944}{{\ttfamily 1803.01944}}].

\bibitem{VOYAGER-SENSITIVITY}
``{LIGO Document T1500293-v13}.''
  \url{https://dcc.ligo.org/LIGO-T1500293/public}.

\bibitem{VOYAGER-bluebird5}
``{LIGO Voyager BlueBird5 sensitivity}.''
  \url{https://dcc.ligo.org/public/0152/G1800986/001/BlueBird5.pdf}.

\bibitem{VOYAGER-upgrade}
``{LIGO Voyager Upgrade: Design Concept}.''
  \url{https://docs.ligo.org/voyager/voyagerwhitepaper/main.pdf}.

\bibitem{Vallisneri:2007ev}
M.~Vallisneri, \emph{{Use and abuse of the Fisher information matrix in the
  assessment of gravitational-wave parameter-estimation prospects}},
  \href{https://doi.org/10.1103/PhysRevD.77.042001}{\emph{Phys. Rev. D}
  {\bfseries 77} (2008) 042001}
  [\href{https://arxiv.org/abs/gr-qc/0703086}{{\ttfamily gr-qc/0703086}}].

\bibitem{Iacovelli:2022mbg}
F.~Iacovelli, M.~Mancarella, S.~Foffa and M.~Maggiore, \emph{{GWFAST: A Fisher
  Information Matrix Python Code for Third-generation Gravitational-wave
  Detectors}}, \href{https://doi.org/10.3847/1538-4365/ac9129}{\emph{Astrophys.
  J. Supp.} {\bfseries 263} (2022) 2}
  [\href{https://arxiv.org/abs/2207.06910}{{\ttfamily 2207.06910}}].

\bibitem{Tamanini:2016zlh}
N.~Tamanini, C.~Caprini, E.~Barausse, A.~Sesana, A.~Klein and A.~Petiteau,
  \emph{{Science with the space-based interferometer eLISA. III: Probing the
  expansion of the Universe using gravitational wave standard sirens}},
  \href{https://doi.org/10.1088/1475-7516/2016/04/002}{\emph{JCAP} {\bfseries
  04} (2016) 002} [\href{https://arxiv.org/abs/1601.07112}{{\ttfamily
  1601.07112}}].

\bibitem{Speri:2020hwc}
L.~Speri, N.~Tamanini, R.R.~Caldwell, J.R.~Gair and B.~Wang, \emph{{Testing the
  Quasar Hubble Diagram with LISA Standard Sirens}},
  \href{https://doi.org/10.1103/PhysRevD.103.083526}{\emph{Phys. Rev. D}
  {\bfseries 103} (2021) 083526}
  [\href{https://arxiv.org/abs/2010.09049}{{\ttfamily 2010.09049}}].

\bibitem{Hirata:2010ba}
C.M.~Hirata, D.E.~Holz and C.~Cutler, \emph{{Reducing the weak lensing noise
  for the gravitational wave Hubble diagram using the non-Gaussianity of the
  magnification distribution}},
  \href{https://doi.org/10.1103/PhysRevD.81.124046}{\emph{Phys. Rev. D}
  {\bfseries 81} (2010) 124046}
  [\href{https://arxiv.org/abs/1004.3988}{{\ttfamily 1004.3988}}].

\bibitem{Kocsis:2005vv}
B.~Kocsis, Z.~Frei, Z.~Haiman and K.~Menou, \emph{{Finding the electromagnetic
  counterparts of cosmological standard sirens}},
  \href{https://doi.org/10.1086/498236}{\emph{Astrophys. J.} {\bfseries 637}
  (2006) 27} [\href{https://arxiv.org/abs/astro-ph/0505394}{{\ttfamily
  astro-ph/0505394}}].

\bibitem{He:2019dhl}
J.-h.~He, \emph{{Accurate method to determine the systematics due to the
  peculiar velocities of galaxies in measuring the Hubble constant from
  gravitational-wave standard sirens}},
  \href{https://doi.org/10.1103/PhysRevD.100.023527}{\emph{Phys. Rev. D}
  {\bfseries 100} (2019) 023527}
  [\href{https://arxiv.org/abs/1903.11254}{{\ttfamily 1903.11254}}].

\bibitem{Meszaros:1995dj}
P.~Meszaros and A.~Meszaros, \emph{{The Brightness distribution of bursting
  sources in relativistic cosmologies}},
  \href{https://doi.org/10.1086/176026}{\emph{Astrophys. J.} {\bfseries 449}
  (1995) 9} [\href{https://arxiv.org/abs/astro-ph/9503087}{{\ttfamily
  astro-ph/9503087}}].

\bibitem{Meszaros:2011zr}
A.~Meszaros, J.~Ripa and F.~Ryde, \emph{{Cosmological effects on the observed
  flux and fluence distributions of gamma-ray bursts: Are the most distant
  bursts in general the faintest ones?}},
  \href{https://doi.org/10.1051/0004-6361/201014918}{\emph{Astron. Astrophys.}
  {\bfseries 529} (2011) A55}
  [\href{https://arxiv.org/abs/1101.5040}{{\ttfamily 1101.5040}}].

\bibitem{Band:2002te}
D.L.~Band, \emph{{Comparison of the gamma-ray burst sensitivity of different
  detectors}}, \href{https://doi.org/10.1086/374242}{\emph{Astrophys. J.}
  {\bfseries 588} (2003) 945}
  [\href{https://arxiv.org/abs/astro-ph/0212452}{{\ttfamily
  astro-ph/0212452}}].

\bibitem{Wanderman:2014eza}
D.~Wanderman and T.~Piran, \emph{{The rate, luminosity function and time delay
  of non-Collapsar short GRBs}},
  \href{https://doi.org/10.1093/mnras/stv123}{\emph{Mon. Not. Roy. Astron.
  Soc.} {\bfseries 448} (2015) 3026}
  [\href{https://arxiv.org/abs/1405.5878}{{\ttfamily 1405.5878}}].

\bibitem{Chen:2020zoq}
H.-Y.~Chen, P.S.~Cowperthwaite, B.D.~Metzger and E.~Berger, \emph{{A Program
  for Multimessenger Standard Siren Cosmology in the Era of LIGO A+, Rubin
  Observatory, and Beyond}},
  \href{https://doi.org/10.3847/2041-8213/abdab0}{\emph{Astrophys. J. Lett.}
  {\bfseries 908} (2021) L4}
  [\href{https://arxiv.org/abs/2011.01211}{{\ttfamily 2011.01211}}].

\bibitem{Ronchini:2022gwk}
S.~Ronchini, M.~Branchesi, G.~Oganesyan, B.~Banerjee, U.~Dupletsa, G.~Ghirlanda
  et~al., \emph{{Perspectives for multimessenger astronomy with the next
  generation of gravitational-wave detectors and high-energy satellites}},
  \href{https://doi.org/10.1051/0004-6361/202243705}{\emph{Astron. Astrophys.}
  {\bfseries 665} (2022) A97}
  [\href{https://arxiv.org/abs/2204.01746}{{\ttfamily 2204.01746}}].

\bibitem{Foreman-Mackey:2012any}
D.~Foreman-Mackey, D.W.~Hogg, D.~Lang and J.~Goodman, \emph{{emcee: The MCMC
  Hammer}}, \href{https://doi.org/10.1086/670067}{\emph{Publ. Astron. Soc.
  Pac.} {\bfseries 125} (2013) 306}
  [\href{https://arxiv.org/abs/1202.3665}{{\ttfamily 1202.3665}}].

\bibitem{Barausse:2012fy}
E.~Barausse, \emph{{The evolution of massive black holes and their spins in
  their galactic hosts}},
  \href{https://doi.org/10.1111/j.1365-2966.2012.21057.x}{\emph{Mon. Not. Roy.
  Astron. Soc.} {\bfseries 423} (2012) 2533}
  [\href{https://arxiv.org/abs/1201.5888}{{\ttfamily 1201.5888}}].

\bibitem{Chen:2018dbv}
L.~Chen, Q.-G.~Huang and K.~Wang, \emph{{Distance Priors from Planck Final
  Release}}, \href{https://doi.org/10.1088/1475-7516/2019/02/028}{\emph{JCAP}
  {\bfseries 02} (2019) 028}
  [\href{https://arxiv.org/abs/1808.05724}{{\ttfamily 1808.05724}}].

\bibitem{Brout:2022vxf}
D.~Brout et~al., \emph{{The Pantheon+ Analysis: Cosmological Constraints}},
  \href{https://doi.org/10.3847/1538-4357/ac8e04}{\emph{Astrophys. J.}
  {\bfseries 938} (2022) 110}
  [\href{https://arxiv.org/abs/2202.04077}{{\ttfamily 2202.04077}}].

\end{thebibliography}\endgroup
\bibliographystyle{JHEP}

\end{document}